\documentclass[usenatbib,fleqn,a4paper]{mnras}
\usepackage{graphicx}
\usepackage{times}
\usepackage{color}
\usepackage{amsmath}

\usepackage{geometry}
\title[Insignificance of mini-haloes to reionisation]
{Feedback-regulated star formation and escape of LyC photons from mini-haloes during reionisation}
\author[Taysun Kimm et al.]{  
\parbox[t]{\textwidth}{
Taysun Kimm$^{1}$\thanks{e-mail: tkimm@ast.cam.ac.uk}, 
Harley Katz$^{1}$, 
Martin Haehnelt$^{1}$, 
Joakim Rosdahl$^{2}$, 
Julien Devriendt$^{3,4}$,
Adrianne Slyz$^3$} 
\vspace*{6pt} \\
$^1$ Kavli Institute for Cosmology and Institute of Astronomy, Madingley Road, Cambridge CB3 0HA, UK\\
$^2$ Leiden Observatory, Leiden University, P.O. Box 9513, 2300 RA, Leiden, The Netherlands\\
$^3$ Astrophysics, University of Oxford, Denys Wilkinson Building, Keble Road, Oxford OX1 3RH, UK\\
$^4$ Observatoire de Lyon, UMR 5574, 9 avenue Charles Andre, F-69561 Saint Genis Laval, France
}

\begin{document}
\maketitle

\newcommand{\fbar}{\mbox{$f_{\rm bar}$}}
\newcommand{\mbar}{\mbox{$m_{\rm bar}$}}
\newcommand{\nH}{\mbox{$n_{\rm H}$}}
\newcommand{\kms}{\mbox{${\rm km\,s^{-1}}$}}
\newcommand{\msun}{\mbox{$\rm M_\odot$}}
\newcommand{\Zsun}{\mbox{$\rm Z_\odot$}}
\newcommand{\msunyr}{\mbox{$\rm M_\odot\,{\rm yr^{-1}}$}}
\newcommand{\mvir}{\mbox{$M_{\rm vir}$}}
\newcommand{\mgas}{\mbox{$M_{\rm gas}$}}
\newcommand{\mstar}{\mbox{$M_{\rm star}$}}
\newcommand{\sfr}{\mbox{$\dot{M}_{\rm star}$}}
\newcommand{\mhalo}{\mbox{$M_{\rm halo}$}}
\newcommand{\rvir}{\mbox{$R_{\rm vir}$}}
\newcommand{\mn}{\mbox{{\sc \small Horizon}-MareNostrum}}
\newcommand{\nut}{\mbox{{\sc \small Nut}}}
\newcommand{\ramses}{\mbox{{\sc \small Ramses}}}
\newcommand{\nth}{\mbox{$n_{\rm SF}$}}
\newcommand{\cmq}{\mbox{${\rm cm^{-3}}$}}
\newcommand{\fesc}{\mbox{${\rm f_{\rm esc}}$}}
\newcommand{\lya}{\mbox{Ly$\alpha$}}



\begin{abstract}
Reionisation in the early Universe is likely driven by dwarf galaxies. 
Using cosmological radiation-hydrodynamic simulations, we study star formation 
and the escape of Lyman continuum (LyC) photons from mini-haloes with $\mhalo \la 10^8\,\msun$. 
Our simulations include a new thermo-turbulent star formation model, non-equilibrium 
chemistry, and relevant stellar feedback processes (photoionisation by young massive stars, 
radiation pressure, and mechanical supernova explosions). 
We find that feedback reduces star formation very efficiently in mini-haloes,
resulting in the stellar mass consistent with the slope and normalisation reported 
in Kimm \& Cen (2014) and the empirical stellar mass-to-halo mass relation derived in the local Universe.
Because star formation is stochastic and dominated by a few gas clumps, 
the escape fraction in mini-haloes is generally determined by radiation feedback 
(heating due to photo-ionisation), rather than supernova explosions. 
We also find that the photon number-weighted mean escape 
fraction in mini-haloes is higher  ($\sim20$--$40\%$) than that in atomic-cooling haloes, 
although the instantaneous fraction in individual haloes varies significantly. The escape fraction 
from Pop III stars is found to be significant ($\ga10\%$) only when the mass is greater than 
$\sim$100\,\msun. Based on simple analytic calculations, we show that LyC photons 
from mini-haloes are, despite their high escape fractions, of minor importance for reionisation 
due to inefficient star formation. We confirm previous claims that stars in atomic-cooling haloes with 
masses  $10^8\,\msun\la\mhalo \la 10^{11}\,\msun$ are likely to be 
the most important source of reionisation.
\end{abstract}


\voffset=-0.6in
\hoffset=0.2in

\section{Introduction}

Observations of Lyman $\alpha$ opacities in the spectra of quasi-stellar objects (QSOs) 
at high redshift have shown unambiguously that the Universe becomes nearly 
transparent to LyC photons ($\lambda \le 912 \AA$) at $z\sim6$ \citep{fan01,becker01,fan06,mcgreer15}.
Several candidates are identified as a potential source of reionisation,
including dwarf galaxies \citep[e.g.,][]{couchman86,madau99}, 
active galactic nuclei \citep[e.g.,][]{shapiro87,haiman98}, 
accretion shock \citep{dopita11}, globular clusters \citep{ricotti02,katz13,katz14}, and 
X-rays from accreting stellar-mass black holes \citep[e.g.,][]{madau04,ricotti04,mirabel11}.
Many studies agree that the primary source of reionisation is likely to be massive stars 
in dwarf galaxies \citep[][c.f., \citealt{madau15}]{haehnelt01,cowie09,fontanot14,madau16}; however, the timescale over which reionisation occurred and the mass range of haloes which provided 
the majority of the ionising photons 
are issues that remain unresolved \citep[e.g.,][]{bolton07,kuhlen12,ahn12,wise14}.

The two critical ingredients for reionisation are 
star formation and escaping LyC photons. The former describes 
how many LyC photons are available from massive stars, while the latter 
determines what fraction are actually used to ionise the intergalactic medium (IGM).
Unsurprisingly, the prediction of both quantities is very challenging, 
as galaxy evolution involves highly non-linear processes, such as 
the interaction between the ISM and feedback from stars. 
For this reason, numerical studies often report discrepant results 
on the escape fraction. An early attempt by \citet{gnedin08} suggested that the escape 
fraction roughly increases with halo mass in the range 
$10^{10}$--$10^{12}\,\msun$, because stellar disks  in lower mass haloes tend to be 
embedded in gaseous disks in their simulations.
\citet{wise09} also found a positive correlation between the escape fraction 
and halo mass in the range $10^{6}$--$10^{10}\,\msun$,
but as \citet{wise14} pointed out, this may have been caused by 
the initial strong starburst due to the absence of cooling while constructing 
the initial conditions of the simulations.
On the contrary, by post-processing hydrodynamics simulations with strong 
stellar feedback, \citet{razoumov10} concluded that low-mass haloes ($\sim10^8\msun$) 
show a higher escape fraction of $\sim100\%$, while less than 10\% of LyC photons 
escape from the massive haloes with mass $10^{11}\,\msun$. A negative dependence on halo
mass in the atomic-cooling regime is also found in other large cosmological 
simulations based on post-processing \citep{yajima11,paardekooper13}.
Large-scale simulations often predict very high escape fractions 
of $\fesc\ga 50\%$ in atomic cooling haloes, but these conclusions may be 
subject to numerical resolution \citep{ma15,paardekooper15} and 
how the escape fraction is measured (i.e. whether or not it is photon number-weighted \citep{kimm14}).
\citet{paardekooper15} pointed out that significant absorption of LyC photons 
occurs on giant molecular cloud scales (i.e. 10 pc). 
Recent theoretical work based on effective feedback with high numerical resolution 
($\la {\rm 10\,pc}$) suggests that, on average, only $\sim10\%$ of LyC photons escape from their host 
haloes with the mass range $10^8\,\msun \la \mhalo \la 10^{11}\,\msun$ 
\citep{kimm14,ma15,xu16}.
The only exception to this relatively low escape fraction found in numerical simulations 
is mini-haloes where $40$--$60\,\%$ of  the ionising photons
escape the galaxies and contribute to the ionisation of the IGM \citep{wise14,xu16}.

In observations, the leakage of LyC photons is measured via 
the relative flux density ratio ($F_{\rm UV}/F_{\rm LyC}$) between the ionising part of the 
spectrum at 900 $\AA$ and the non-ionising part at $1500\AA$ 
\citep[][]{steidel01}. Once absorption due to the IGM is corrected 
\citep[e.g.,][]{inoue14}, one can estimate the absolute escape fraction assuming a 
ratio of the intrinsic luminosity at 900 $\AA$ and 1500 $\AA$ appropriate for 
the observed multi-band photometric data.  
The detection of LyC photons in the local Universe is limited to starburst galaxies \citep{leitet11,leitet13,borthakur14,leitherer16}, where generally small escape fractions 
($\la 5 \%$) are observed. Star-forming galaxies at $z\sim1$ with LyC detection also show 
low escape fractions of a few percent \citep{siana07,siana10,bridge10,rutkowski16}.
Efficient LyC leakers ($\fesc\ga10\%$) seem to be more common 
at higher redshift \citep[$z\ga3$, e.g.,][]{reddy16}, but only a handful of cases 
are confirmed as robust detections which are not affegacted by contamination 
due to low-redshift interlopers along the line of sight 
\citep[][]{mostardi15,shapley16,leethochawalit16}.
The average, relative escape fraction in nearly all observations is  
found to be very small, even at high redshift \citep[$f_{\rm esc}^{\rm rel}\la 2\%$,][]{grazian16,vanzella10,boutsia11,siana15,mostardi15}, 
and appears to be in tension with the $\fesc\sim10\%$
needed to reconcile the observed luminosity of high-redshift galaxies
with observational constraints on the evolution of  the average neutral hydrogen fraction.
However, it should be noted that these estimates  mostly focus on small 
galaxies of mass $\mstar\le10^8\,\msun$ or $M_{\rm UV}\ga -18$ at $z\ge6$, 
whereas observed samples are biased towards bright galaxies 
\citep[$M_{\rm UV}\la-20$, e.g.,][]{grazian16}.
Because star formation in small galaxies is more bursty than in bright 
galaxies observed at lower redshift \citep[e.g.,][]{speagle14}, 
it is conceivable that the star-forming clouds are disrupted more efficiently in simulated galaxies,
resulting in higher escape fractions \citep[e.g.,][]{kimm14,cen15}.
Moreover, since the simulated galaxies are more metal-poor than the observed bright galaxies,
they are likely less affected by dust compared to observed galaxies \citep[e.g.,][]{izotov16}.
Finally, as pointed out by \citet{cen15}, individual measurements of the escape fraction
may underestimate the three dimensional escape fraction, especially when 
the escape fraction is small.

Unlike the observed LyC flux that conveys information
about the instantaneous escape fraction, 
the Thompson electron optical depth ($\tau_e$) derived from the polarisation signal of  
cosmic microwave background (CMB) photons provides a useful measure of 
how extended reionisation was in the early Universe.
The analysis of the nine-year Wilikinson Microwave Anisotropy Probe (WMAP9) 
observations suggested a high electron optical depth of $\tau_e=0.089 \pm 0.014$ \citep{hinshaw13}, 
indicating that ionised hydrogen (HII) bubbles are likely to have grown relatively early.
However, the observed number density of bright galaxies in the 
ultraviolet (UV) ($M_{\rm UV}\la-17$) is unable to explain such a high 
$\tau_e$  \citep[e.g.,][]{bunker10,finkelstein10,bouwens12}.
By taking a parametric form of the UV luminosity density, motivated by observations of 
the Hubble Ultra Deep Field, \citet{robertson13} showed that the inclusion of small 
dwarf galaxies with $-17 \le M_{\rm UV} \le -13$ can increase $\tau_e$ to a 
higher value of $0.07$, provided that $20\%$ of LyC photons 
escape from the dark matter haloes. \citet{wise14} claim that 
mini-haloes of mass $\mhalo\le10^8\,\msun$, corresponding to $M_{\rm UV} \ga -13$, 
may be able to provide a large number of LyC photons to the IGM 
as LyC photons escape freely from their host halo. Because the mini-haloes emerge first 
and they are abundant in the early Universe ($z\ge15$), the authors find that the resulting 
$\tau_e\approx0.09$ can easily accommodate the WMAP9 analysis, demonstrating the potential
importance of mini-haloes to reionisation of the Universe \citep[see also][]{ahn12}.

However, a more accurate modelling of dust emission in our Galaxy 
\citep{planck-collaboration14} and the use of the low frequency instrument on the 
Planck satellite lead to a decrease in the optical depth to $\tau_e=0.066\pm0.016$ 
\citep{planck-collaboration15}. The latest results utilising the high frequency 
instrument to measure the low-multipole polarisation signal point to a 
possibility of an even lower value of $\tau_e=0.055\pm0.009$ \citep{planck-collaboration16}. 
Furthermore, recent findings of  significant \lya\ opacity fluctuations on large scales in absorption 
spectra of $5\la z\la 6 $ QSOs \citep{becker15} and the observed rapid evolution of \lya\ emitters at $z>6$  \citep{ono12,caruana14,schenker14,pentericci14,tilvi14,matthee15}
suggest that reionisation may have ended later than previously thought \citep[e.g.,][c.f., \citealt{haardt12,mesinger15}]{chardin15,choudhury15,davies16}.
If the contribution from mini-haloes were  important for reionisation, this may potentially be in tension with the 
reduced $\tau_e$ measurement and the long  \lya\ troughs still observed at $z\sim 5.6$ \citep{becker15}. 
Therefore, in this study, we revisit the importance of mini-haloes 
and assess their role in reionisation using state-of-the-art numerical simulations.

This paper is organised as follows. In Section 2, we describe the physical ingredients 
used in our cosmological, radiation-hydrodynamic simulations. The measurements
of the escape fraction and star formation in the simulations are presented in Section 3. 
Section 4 discusses the mechanisms responsible for the escape of LyC photons
and whether or not the ionising radiation from mini-haloes is crucial to reionisation of the Universe.
We summarise our findings in Section 5.

\section{Simulation}

We use {\sc ramses-rt}, a radiation hydrodynamics code with adaptive mesh refinement, 
\citep{teyssier02,rosdahl13,rosdahl15a}, to study reionisation 
due to starlight in mini-haloes with $10^6 \la \mhalo/\msun \la 10^{8}$.
The cosmological initial conditions are generated using {\sc music} \citep{hahn11}, 
with the cosmological parameters
($\Omega_{\rm m}=0.288$, 
$\Omega_{\rm \Lambda}=0.712$, 
$\Omega_{\rm b}=0.045$, 
$H_0=69.33\,{\rm km\,s^{-1}\,Mpc^{-1}}$, 
$n_s=0.971$, and 
$\sigma_8=0.830$) consistent 
with the WMAP9 results \citep{hinshaw13}.
We first run dark matter only simulations with volume $(2\,{\rm Mpc}/h)^3$, and 
identify nine regions hosting a halo of mass $\approx 10^8\,\msun$ at $7\le z \la 11$.
The initial conditions for the zoom-in regions are then generated with a higher dark matter 
resolution of 90 \msun\ to resolve each halo with more than 10,000 dark matter particles.
We ensure that the haloes are not contaminated by coarse dark matter particles.

We solve the Euler equations using a HLLC scheme \citep{toro94}, with the typical 
courant number of 0.7. The Poisson equation is solved using a multigrid method  \citep{guillet11}. For the transport of multiple photon groups, {\sc ramses-rt} uses a 
moment-based method with M1 closure for the Eddington tensor \citep[][see also \citealt{aubert08}]{rosdahl13,rosdahl15a}. 
We adopt a GLF scheme to solve the advection of the photon fluids.
Because the hydrodynamics is fully coupled to the radiation, 
the computational time step is usually determined by the speed of light.
Since we are interested in the escaping flux, which is a conserved quantity,
we use a reduced speed of light approximation ($\tilde{c} = 3\times10^{-3} c$) 
to keep the computational cost low, where $c$ is the full speed of light.

Each zoom-in simulation is covered with $128^3$ root cells,
and we allow for further refinement of the computational grid to achieve a maximum 
physical resolution of 0.7 pc. To do so, we adopt two different refinement criteria. 
First, a cell is refined if the total baryonic plus dark matter inside each cell exceeds 
8 times the mass of a dark matter particle (i.e. 720 \,\msun). 
Second, we enforce that the thermal Jeans length is resolved by at least 32 cells 
until it reaches the maximum resolution. Although the use of the latter condition is 
computationally expensive, it makes our simulations more robust than previous 
simulations, as the turbulent properties of gas can be more accurately captured \citep{federrath11,turk12,meece14}.

We identify dark matter haloes with the AdaptaHop algorithm \citep{tweed09,aubert04}. 
The centre of a dark matter halo is chosen as the centre of mass of the star particles in the halo.
If a halo is devoid of stars, we use the densest location of the halo. 
The virial mass and radius of a halo is computed such that the mean density 
within the virial sphere is equivalent to $\Delta_{\rm crit} \rho_{\rm crit}$,
where $\rho_{\rm crit}$ is the critical density of the universe ($3H(z)^2/8\pi G$),  
$\Delta_{\rm crit}=18\pi^2 + 82 x - 39 x^2$ is the virial overdensity \citep{bryan98}, 
$x\equiv\Omega_{\rm m} /\left(\Omega_{\rm m} + a^3 \Omega_{\rm \Lambda}\right) -1$,
$G$ is the gravitational constant, and $H(z)$ is the Hubble constant at some redshift $z$.

\begin{table}
   \caption{Summary of simulation parameters.}
   \centering
   \begin{tabular}{lll}
   \hline
  Parameter & Value & Description  \\
     \hline
    $L_{\rm box}$ & $2\,{\rm Mpc} \, h^{-1}$ & Simulation box size \\
    $\Delta x_{\rm min}$ & 0.7 pc & physical size of the finest cell \\
     $m_{\rm DM}$ & 90 \msun & mass resolution of DM particles \\
     $m_{\rm star}$ & 91 \msun & mass resolution of Pop II star particles\\
     $\lambda_{\rm J}/\Delta x$ & 32 & Jeans length criterion for refinement \\
     $\epsilon_{\rm ff}$ & FK12 & SF efficiency per free-fall time \\
     $N_{\rm halo}$ & 9 & number of zoom-in haloes\\
     IMF & \citet{kroupa01} &  (i.e., 1 SN per 91 \msun)\\
       \hline
   \end{tabular}
   \label{tab:sim}
\end{table}

\subsection{Star formation}

Star formation is modelled based on a Schmidt law \citep{schmidt59},
\begin{equation}
\frac{{\rm d} \rho_{\rm star}}{{\rm d} t} = \epsilon_{\rm ff} \frac{\rho_{\rm gas}}{t_{\rm ff}},
\end{equation}
where $\rho_{\rm gas}$ is the density of gas, and $t_{\rm ff}=\sqrt{3\pi/32 G \rho_{\rm gas}}$ 
is the free-fall time. The main parameter characterising star formation  is the star formation efficiency 
per free-fall time ($\epsilon_{\rm ff}$). Local observations find that the efficiency is 
only a few percent when averaged over galactic scales \citep[e.g.,][]{kennicutt98}. However, 
recent findings from small-scale numerical simulations suggest that 
$\epsilon_{\rm ff}$ depends on physical properties of the ISM \citep{padoan11,federrath12}.
Motivated by this, we adopt a thermo-turbulent star formation model in which 
$\epsilon_{\rm ff}$ is determined on a cell-by-cell basis 
(Devriendt et al. 2016, {\sl in prep}). The details of the model will be presented elsewhere, and here, we briefly describe the basic idea for completeness.

The most fundamental assumption in the thermo-turbulent model is that 
the probability distribution function (PDF) of the density of a star-forming cloud is well described by a log-normal distribution. 
By integrating the gas mass from some critical density above which gas can collapse 
($\rho = \rho(s_{\rm crit})$) to infinity ($\rho = \infty$) per individual free-fall time, 
$\epsilon_{\rm ff}$ can be estimated as \citep[e.g.,][]{federrath12}
\begin{equation}
\epsilon_{\rm ff} = \frac{\epsilon_{\rm ecc}}{2 \phi_t} \exp\left(\frac{3}{8}\sigma_s^2 \right) \left[ 1 + {\rm erf} \left(\frac{\sigma_s^2 - s_{\rm crit}}{\sqrt{2\sigma_s^2}}\right)\right],
\label{eq:sfe}
\end{equation}
where $\sigma_s^2 = \ln \left(1 + b^2 \mathcal{M}^2 \right)$ is the standard deviation 
of the logarithmic density contrast ($s \equiv \ln \left(\rho/\rho_0\right)$), $\rho_0$ is the mean 
density of gas, $b$ is a parameter that depends on the mode of turbulence driving, 
$\epsilon_{\rm ecc}\approx0.5$ is the maximum fraction of gas that can accrete onto stars 
without being blown away by proto-stellar jets and outflows, 
$\phi_t\approx0.57$ is a factor that accounts for 
the uncertainties in the estimation of a free-fall time of individual clouds,
and $\mathcal{M}$ is the sonic Mach number. We assume a mixture of solenoidal and 
compressive modes for turbulence ($b\approx0.4$). An important quantity in Equation~\ref{eq:sfe} is 
the critical density ($s_{\rm crit}$), which may be regarded as the minimum density above 
which gas in the post-shock regions of a cloud is magnetically supercritical and 
thus can collapse \citep{krumholz05,padoan11,hennebelle11}.
Numerical simulations suggest that $s_{\rm crit}$ may be approximated as \citep{padoan11,federrath12} 
\begin{equation}
s_{\rm crit} = {\rm ln} \left(0.067\theta^{-2}\,\alpha_{\rm vir} \,\mathcal{M}^2\right),
\end{equation}
where $\theta$ is a numerical factor of order unity that encapsulates the uncertainty in the
post-shock thickness with respect to the cloud size, and we adopt $\theta=0.33$ 
that gives a best fit to the results of \citet{federrath12}.
Here $\alpha_{\rm vir}\equiv 2 E_{\rm kin}/ |E_{\rm grav}$ is the virial parameter of a cloud, which we take to be 
$\alpha_{\rm 0} \approx 5(\sigma_{\rm gas}^2 + c_s^2)/(\pi \rho_{\rm gas} G \Delta x^2)$,
where $\Delta x$ is the size of a computational cell, $c_s$ is the gas sound speed,
and $\sigma_{\rm gas}$ is the turbulent gas velocity, 
which is the tracer of the velocity gradient.
Note that the resulting $\epsilon_{\rm ff}$ can  be larger 
than 1 (Figure~1 of \citealt{federrath12}) if the sonic Mach number 
is very high ($\mathcal{M}\ga10$) in tightly gravitationally bound 
regions ($\alpha_{\rm vir} \la 0.1$).
In practice, we find that clouds with such conditions are extremely rare in our simulations, and 
$\epsilon_{\rm ff}$ typically ranges from 5\% to 20\% when star particles are created.

This thermo-turbulent model allows for star formation only if the thermal plus turbulent 
pressure is not strong enough to prevent the gravitational collapse of a gas cloud. 
This may be characterised by the turbulent Jeans length \citep{bonazzola87,federrath12} 
\begin{equation}
\lambda_{\rm J,turb} = \frac{\pi \sigma_{\rm gas}^2 + \sqrt{36 \pi c_s^2 G \Delta x^2 \rho_{\rm gas} + \pi^2\sigma_{\rm gas}^4 }}{6 G \rho_{\rm gas} \Delta x}.
\end{equation}
In order for gas to be gravitationally unstable,  the Jeans length needs 
to be smaller than  the size of a computational cell ($ \lambda_{\rm J,turb} \le \Delta$). 
Note that star formation occurs only in the maximally refined cells,
because our refinement strategy enforces the thermal Jeans length to be resolved by 32 cells 
until it reaches the maximum level of refinement.

Once a potential site for star formation is identified, we estimate $\epsilon_{\rm ff}$ 
using Equation~(\ref{eq:sfe}) and determine the number of newly formed stars ($N_\star$) 
based on a Poisson distribution 
\begin{equation}
P(N_\star) = \frac{\lambda^{N_\star}}{N_\star!} \exp\left(-\lambda\right),
\end{equation}
with a mean of 
\begin{equation}
\lambda = \epsilon_{\rm ff} \frac{\rho_{\rm gas} \Delta x^3}{m_{\rm \star, min}} \left(\frac{\Delta t_{\rm sim}}{t_{\rm ff} }\right),
\end{equation}
where $m_{\rm \star, min}$ is the minimum mass of a star particle, 
and $\Delta t_{\rm sim}$ is the simulation integration time step.
We adopt $m_{\rm \star,min}=91\,\msun$ for Pop II stars\footnote{We discuss the possible impact of  
the IMF sampling, the neglect of runaway stars, and the uncertainties in stellar evolutionary models on the determination of the escape fractions in Section~\ref{sec:caveats}.}, which would host a single SN 
for a Kroupa initial mass function \citep{kroupa01}.
Of the stellar mass, 21 percent is returned to the surrounding medium as a result of 
SN explosions, and 1 percent is assumed to be newly synthesised to metals 
(i.e. a metal yield of 0.01). 

The formation of Pop III stars is included following \citet{wise12a}.
We adopt a Salpeter-like IMF for masses above the characteristic mass ($M_{\rm char}$),
while the formation of low-mass Pop III stars is assumed to be inefficient, 
\begin{equation}
\frac{dN}{d\log M} \propto M^{-1.3} \exp{\left[ - \left( \frac{M_{\rm char}}{M}\right)^{1.6}\right]},
\label{eq:sf_pop3}
\end{equation}
where $N$ is the number of Pop III stars per logarithmic mass bin.
The precise determination of the characteristic mass is a matter of debate. 
Early studies suggested that the mass of protostellar clumps is $\sim 100\,\msun$ \citep{abel02,bromm02b,yoshida06}. 
Later, several groups point out that gas clumps may be fragmented further reducing the 
characteristic mass to $\sim 40\,\msun$ \citep{turk09,greif12}
However, recent radiation-hydrodynamics simulations report that several tens to a thousand solar masses of 
gas may collapse to form a Pop III star \citep[][c.f., \citealt{lee16}]{hirano14,hosokawa15}.
In this work, we adopt $M_{\rm char}=100\,\msun$, consistent with the most recent simulations.

We assume that Pop III stars form only in a region where the gas metallicity is below $10^{-6}\,\Zsun$.
This means that at least one Pop III star will form in a dark matter halo during the initial gas collapse 
due to radiative cooling by molecular hydrogen. In principle, the external pollution by neighbouring 
haloes can suppress the formation of Pop III stars \citep[e.g.][]{smith15}, but our simulated haloes 
are chosen to reside in an isolated environment and thus are not affected by neighbours. 
Note that more than one Pop III star can form in each halo if the first Pop III star does not explode 
and enrich the IGM/ISM or if a pristine gas cloud is accreted onto a dark matter halo through halo mergers. 

\subsection{Stellar feedback}
Modelling the feedback from stars is essential to predict the escape of LyC photons in dwarf galaxies.
In order for LyC photons to leave their host dark matter halo, feedback should clear away low-density
channels or entirely blow out the birth clouds. Otherwise, the photons will simply 
be absorbed by  neutral hydrogen inside of the halo. We include three different 
types of feedback (photoionisation, radiation pressure from the absorption of UV and 
IR photons, and Type II supernova feedback) in our simulations. 

\subsubsection{Radiation feedback}

Young, massive stars emit large amounts of ionising photons, which drive winds 
through various processes. Because the absorption cross section of neutral hydrogen 
is so large ($\sigma_{\rm abs}\sim 6\times10^{-18}\,{\rm cm^2}$), the presence of a small 
amount of hydrogen makes the ISM optically thick to photons with $E>13.6\,{\rm eV}$. 
When the ISM is fully ionised, dust becomes the next most efficient absorber, as its opacity 
in the UV wavelengths is large as well 
($\kappa_{\rm abs}\sim 1000 \, {\rm cm^2 / g}$). 

Of the several radiation feedback processes, photoionisation is probably the most 
important mechanism that governs the dynamics of a giant molecular cloud \citep[GMC, ][]{lopez14,dale14,rosdahl15a}.
LyC photons can ionise hydrogen, which heat the  gas to $T\approx2\times10^4\,{\rm K}$. 
This creates an over-pressurised HII bubble that lowers the density of the ambient 
medium and drives winds with velocities up to 10 ${\rm km\,s^{-1}}$ \citep[e.g.][]{krumholz07b,walch12,dale14}. 
To capture the dynamics of the HII region, the Stromgren sphere radius ($r_{\rm S}$) 
should be resolved.
\begin{align}
r_{\rm S} & = \left( \frac{3 \dot{N}_{\rm p}}{4 \pi \alpha_{\rm B} n_{\rm H}^2}\right)^{1/3} \nonumber \\
 & \approx 1.2\,{\rm pc} \left( \frac{m_{\rm star}}{10^3 \msun} \right)^{1/3} \left( \frac{n_{\rm H}}{10^3\,{\rm cm^{-3}}} \right)^{-2/3},
\end{align}
where $\dot{N}_p$ is the production rate of ionising photons, and
$\alpha_{\rm B}=2.6\times10^{-13}\,{\rm cm^{3}s^{-1}}$ is the case B recombination rate 
coefficient at ${\rm T=10^4 K}$. For the latter equality, we use 
$N_p=5\times10^{46}\,{\rm s^{-1}}$ per $1\msun$. Note that this scale describes the 
maximum distance within which recombination is balanced by ionisation assuming 
that photo-heating does not affect the dynamics of the ISM.

When ionising photons are absorbed by the neutral ISM, their momentum is  
transferred to the medium \citep[e.g.,][]{haehnelt95} at a rate
\begin{equation}
\dot{\bf p}_{\gamma} = \sum_{i}^{\rm groups} \frac{{\bf F}_i}{c} \left( \kappa_i + \sum_{j}^{ {\rm H I}, {\rm He I}, {\rm He II}, {\rm LW}} \sigma_{ij}n_j\right),
\end{equation}
where $\bf{F}_i$ is the photon flux ($\rm erg\, cm^{-2} s^{-1}$) for $i$-th photon group, 
$\kappa$ is the dust opacity ($\rm cm^2 g^{-1}$), $\sigma$ is the photoionisation 
cross-section ($\rm cm^2$), and $n_j$ is the number density ($\rm cm^{-3}$) of the 
ion species $j$. For a Kroupa IMF, direct radiation pressure from ionising radiation 
($\lambda \le 912\AA$) can impart  momentum of up to $\sim 40 \,\kms\,\msun$ 
if we integrate the number of ionising photons from a simple stellar population of 
$1\,\msun$ until 50 Myr \citep{leitherer99}. The absorption of non-ionising 
UV and optical photons by dust can further increase  the total momentum 
input to $\sim 190 \,\kms\ \,\msun$ within  50 Myr. 

We also take into account radiation pressure by trapped infrared (IR) photons. 
IR photons are generated when UV and optical photons are absorbed by dust 
or when molecular hydrogen is fluorescently excited by absorption of Lyman-Werner photons  
and radiatively de-excited through forbidden rotational-vibrational transitions 
(see the Chemistry section). We assume that these IR photons are efficiently trapped only if the optical depth over the cell width by dust is high. 
The resulting trapped IR photon energy in each cell is modelled as \citep{rosdahl15a}
\begin{equation}
E_{\rm IR, trapped} = f_{\rm trapped} E_{\rm IR} =\exp \left( - \frac{2}{3\,\tau_{\rm d}}\right)\,  E_{\rm IR} 
\end{equation}
where $\tau_{\rm d} = \kappa_{\rm sc} \rho_{\rm gas} \Delta x$
and $\kappa_{\rm sc}\sim 5\, (Z_{\rm gas}/Z_\odot)\,{\rm cm^2/g}$ is the scattering 
cross-section by dust \citep{semenov03}.
This trapped IR radiation is then included as a non-thermal pressure term in the momentum equation, as
\begin{equation}
\frac{\partial \rho \bf{v}}{\partial t} + \nabla \cdot ( \rho {\bf v} \otimes {\bf v} + ( P + P_{\rm rad} ) \bf{I} ) = \dot{\bf p}_{\gamma} + \rho \nabla \Phi
\end{equation}
where 
\begin{equation}
P_{\rm rad} = \frac{\tilde{c}}{c}\frac{E_{\rm IR, trapped}}{3}.
\end{equation}
The non-thermal pressure imparted by trapped IR radiation ($P_{\rm rad}$) is also added 
to the energy equation. Note that these trapped IR photons are advected with the gas, 
while the remaining fraction of the IR energy density
\begin{equation}
E_{\rm IR, stream} = \left[ 1 - \exp \left( - \frac{2}{3\,\tau_{\rm d}}\right)\, \right]  E_{\rm IR} 
\end{equation}
 is diffused out to the neighbouring cells and it is re-evaluated whether or not these photons are trapped by dust \citep{rosdahl15a}.
 
 In this paper, we adopt the photon production rates of Pop II stars from \citet{bruzual03} 
assuming a Kroupa IMF. This is done by interpolating the spectral energy distributions 
for a given metallicity and age and by counting the LyC photons from the spectrum of 
each star particle. The lifetime and the photon production rates 
for Pop III stars are taken by fitting the results of \citet{schaerer02}.
 
\subsubsection{Type II SN explosions}

\begin{figure}
   \centering
   \includegraphics[width=8.5cm]{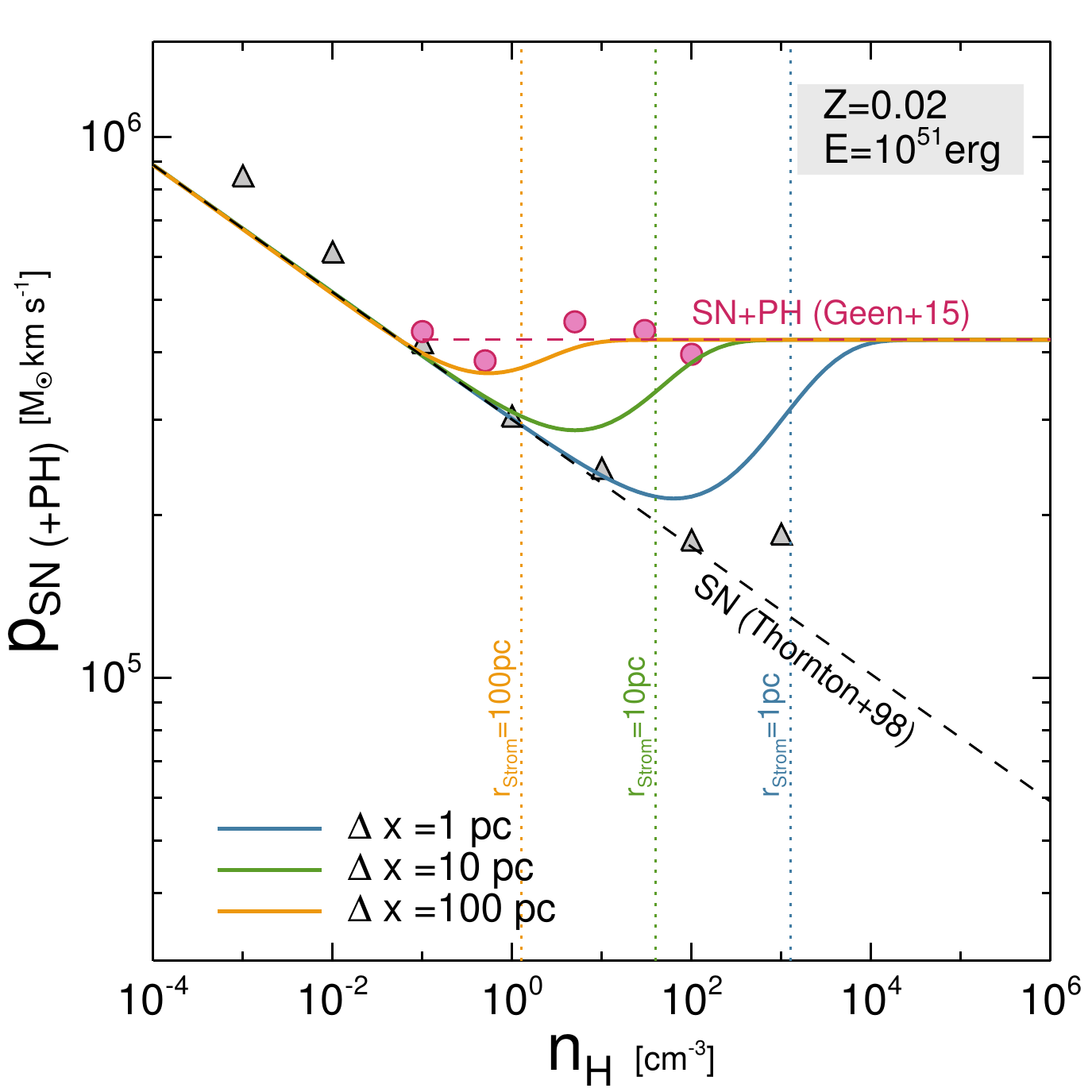} 
   \caption{Final radial momentum from an individual SN explosion 
   in different environments ($p_{\rm SN}$).
   Grey triangles represent the results from 1D hydrodynamic calculations \citep{thornton98}, 
   while violet circles show the final momentum in the presence of ionising radiation 
   from 3D radiation-hydrodynamic simulations \citep{geen15}. 
   The grey and violent dashed lines display a simple fit 
   to these results ($p_{\rm SN}\propto \nH^{-2/17}$, $p_{\rm SN} \sim  {\rm const}$), 
   respectively. The inclusion of photo-heating enhances $p_{\rm SN}$ by reducing 
   the density of the ambient medium before SNe explode and also by imparting 
   momentum from photo-heating and direct radiation pressure. We take this effect 
   into consideration explicitly only if the Stromgren sphere is under-resolved in 
   our simulations. When the Stromgren sphere is resolved, the effect is captured 
   self-consistently in our simulations. The solid lines with different colours illustrate 
   three examples of the momentum that we would impart during the 
   momentum-conserving phase at different resolutions in the presence of the 
   radiation from a star cluster with $M_{\rm star}=10^3\,\msun$, and the dotted lines 
   indicate the corresponding densities at which the Stromgren sphere becomes unresolved. 
   Notice that we inject momentum of the initial ejecta 
   ($p_{\rm SN,ad}=4.5\times10^4\,\msun\,\kms$) if the mass in the neighbouring cells 
   is negligible compared to the ejecta mass (see Equation~\ref{eq:psn}). 
   }
   \label{fig:psn}
\end{figure}

We adopt the mechanical feedback scheme introduced by \citet{kimm14} and \citet{kimm15} 
to model the explosion of massive ($M\ge8\,\msun$) Pop II stars. 
Based on \citet[][see also \citealt{blondin98,kim15,geen15,martizzi15}]{thornton98}, 
this model captures the correct radial momentum input from SN explosions 
at the snowplough phase ($p_{\rm SN,snow}$), 
\begin{equation}
p_{\rm SN, snow} = 3\times10^5\,{\rm km\,s^{-1}\,\msun}\, n_{\rm H}^{-2/17}\, E_{51}^{16/17} \, Z'^{-0.14},
\end{equation}
by imparting momentum according to the stage of the Sedov-Taylor blast wave. 
Here $n_{\rm H}$ is the hydrogen number density in units of ${\rm cm^{-3}}$, 
$E_{51}$ is the explosion energy in units of ${\rm 10^{51}\, erg}$, and 
$Z'={\rm max[0.01,Z/0.02]}$ is the metallicity of gas, normalised to the solar value.

Recently, \citet{geen15} have shown that the final radial momentum from a SN 
can be augmented by including photoionisation from massive stars. 
Because the thermal energy that is liberated during the ionisation process over-pressurises 
and decreases the density of the surroundings into which SNe explode, 
the final momentum from SNe in a medium pre-processed by ionising radiation 
is found to be significantly larger than without it. Most notably, they find that 
the amount of momentum is nearly independent of the background density, 
indicating that more radial momentum should be imparted to the ISM than 
suggested  by  \citet{thornton98}, especially when SNe explode in dense environments.
In principle, this extra momentum should be generated by solving the full radiation hydrodynamics,
but this  requires the simulation to resolve the Stromgren radius.
If young stars are embedded in a cloud denser than $10^3\,{\rm cm^{-3}}$,
the effect of photoionisation is likely to be underestimated with our pc scale resolution. 
In order to circumvent this issue, we adopt a simple fit to the results of \citet{geen15}, 
\begin{equation}
p_{\rm SN+PH} = 4.2\times10^5\,{\rm km\,s^{-1}\,\msun} \,E_{51}^{16/17} \, Z'^{-0.14},
\label{eq:geen}
\end{equation}
if the Stromgren sphere is under-resolved ($\Delta x \gg r_{\rm S}$). 
We adopt the dependence on the SN energy and the metallicity from \citet{thornton98}.
Note that the  values taken from \citet{geen15} are lowered by a factor $1.2^{16/17}$, 
as they use a 20\% larger SN explosion energy ($1.2\times10^{51}\,{\rm erg}$)
compared to other studies \citep{thornton98}. 
The final momentum input during the snowplough phase is then taken from 
the combination of $p_{\rm SN,snow}$ and $p_{\rm SN+PH}$ by
comparing the resolution of a computational cell ($\Delta x$) with the 
Stromgren radius ($r_S$), as
\begin{equation}
p_{\rm SN} = p_{\rm SN, snow} \exp{\left(-\frac{\Delta x}{r_{\rm S}}\right)} + p_{\rm SN+PH} \left(1-\exp{\left[-\frac{\Delta x}{r_{\rm S}}\right]}\right). 
\end{equation}
Figure~\ref{fig:psn} illustrates three examples of the SN momentum that we would inject 
at different resolutions (1, 10, and 100 pc) as a function of density
in the presence of the radiation from a star cluster with $10^3\,\msun$.

Specifically, the model first calculates the mass ratio ($\chi$) 
between the swept-up mass ($M_{\rm swept}$) and the ejecta mass ($M_{\rm ej}$) 
along each $N_{\rm nbor}$ neighbouring cell, as 
\begin{equation}
\chi \equiv dM_{\rm swept} / dM_{\rm ej},
\end{equation}
where 
\begin{equation}
dM_{\rm ej} = (1-\beta_{\rm sn}) M_{\rm ej} / N_{\rm nbor},
and
\end{equation}
\begin{equation}
dM_{\rm swept} = \rho_{\rm nbor} \left(\frac{\Delta x}{2}\right)^3 + \frac{(1-\beta_{\rm sn})\rho_{\rm host}\Delta x^3}{N_{\rm nbor}}  + dM_{\rm ej}.
\end{equation}
Here $\beta_{\rm sn}$ is a parameter that determines what fraction of the gas mass ($M_{\rm ej} + \rho_{\rm host} \Delta x^3$) is re-distributed to the host cell of a SN. In order to distribute the mass evenly to the host and neighbouring cells in the uniformly refined case, we take $\beta_{\rm sn}=4/52$. 
Note that since the maximum number of neighbouring cells is 48 if they are more refined 
than the host cell of a SN (see Figure 15 of \citet{kimm14}), we use $N_{\rm nbor}=48$. 
If the neighbours are not further refined, we simply take the physical properties ($\rho$, $\vec{v}$, $Z$) 
of the neighbours assuming that they are refined. 

We then use the mass ratio ($\chi$) to determine the phase of the Sedov-Blast wave.
To do so, we define the transition mass ratio ($\chi_{\rm tr}$) by 
equating $p_{\rm SN}$ with the radial momentum one would expect during the 
adiabatic phase $p_{\rm ad} = \sqrt{2 \chi M_{\rm ej} f_e E_{\rm SN} }$, where $f_e\sim2/3$ is the 
fraction of energy that is left in the beginning of the snowplough phase, as
\begin{equation}
\chi_{\rm tr} \approx \frac{900\, n_{\rm H}^{-4/17} E_{51}^{-2/17} Z'^{-0.28}}{f_e (M_{\rm ej}/\msun)}.
\end{equation}
If $\chi$ is greater than $\chi_{\rm tr}$, we inject the momentum during the snowplough phase,
whereas the momentum during the adiabatic phase is added to the neighbouring cell, as
\begin{align}
p_{\rm SN} = \left\{ \begin{array}{ll}
     p_{\rm SN, ad}=\sqrt{2 \chi\, M_{\rm ej}\,f_e(\chi) \,E_{\rm SN}}
 & (\chi < \chi_{\rm tr}) \\
    p_{\rm SN}    & (\chi \ge \chi_{\rm tr}) \\ 
   \end{array}
   \right. ,
   \label{eq:psn}
\end{align}
where the fraction of energy left in the SN bubble ($f_e(\chi) \equiv 1-\frac{\chi-1}{3(\chi_{\rm tr}-1)}$) is 
modified to smoothly connect the two regimes.

In order to account for the fact that the lifetime of supernova progenitors varies from 3 Myr to 40 Myr
depending on their mass, we randomly draw the lifetime based on the integrated SN occurrence 
rate from Starburst99 \citep{leitherer99} using the inverse method, as in the MFBmp model from \citet{kimm15}.

\subsubsection{Explosion of Pop III stars}

The explosions of Pop III stars are modelled similarly as Pop II explosions,
but with different energy and metal production rates.
Stars with $40\,\msun \le M_\star \le 120 \,\msun$  and $M_\star \ge 260\,\msun$ are likely to implode 
without releasing energy and metals, while massive stars with $120\,\msun < M_\star < 260\,\msun$ may 
end up as a pair-instability SN \citep{heger03}. The explosions of stars less 
massive than $40\,\msun$ are modelled as either normal Type II SN with 
$10^{51}\,{\rm erg}$ if $11\,\msun\le M_\star \le 20\,\msun$ \citep{woosley95} 
or hypernova if $20\,\msun\le M_\star \le 40\,\msun$ \citep{nomoto06}.
For the explosion energy ($E_{\rm SN,III}$) and returned metal mass ($M_z$), 
we adopt the compilation of \citet{wise12b}, which is based on \citet{woosley95,heger02,nomoto06}, 
as 
\begin{equation}
\frac{E_{\rm SN,III}}{10^{51}\,{\rm erg}}=\left\{
\begin{array}{cc}
 1 &  \left[ 11\le M_\star < 20\right]\\
 \left(-13.714+1.086\,M\right) & \left[ 20\le M_\star \le 40\right]\\ 
 \left(5.0 + 1.304\times(M_{\rm He}-64)\right) &   \left[ 140\le M_\star \le 230\right]\\ 
 0 & ~~~~  {\rm elsewhere}\\ 
\end{array} 
\right. ,
\end{equation}
where $M_{\rm He}=\frac{13}{24}(M_\star-20)$ is the helium core mass, and
\begin{equation}
\frac{M_{\rm z}}{\msun}=\left\{
\begin{array}{cc}
 0.1077 + 0.3383\,(M_\star-11) &  \left[ 11\le M_\star < 20\right]\\
-2.7650 + 0.2794\, M_\star& \left[ 20\le M_\star \le 40\right]\\ 
 (13/24)\,(M_\star-20) &   \left[ 140\le M_\star \le 230\right]\\ 
 0 & ~~~~  {\rm elsewhere}\\ 
\end{array} 
\right. .
\end{equation}
We neglect accretion and feedback from black holes formed 
by the implosion of massive Pop III stars.

\begin{table}
   \caption{Properties of photon groups.}
   \centering
   \begin{tabular}{lcccl}
   \hline
  Photon & $\epsilon_0$ & $\epsilon_1$ & $\kappa$  & Main function \\
  group           &      [eV]          &   [eV] &   [$\rm cm^2/g$]   & \\
     \hline
   IR & 0.1 & 1.0 & 5 & Radiation pressure (RP)\\
   Optical & 1.0 & 5.6 & $10^3$ & Direct RP\\
   FUV & 5.6 & 11.2 & $10^3$ & Photoelectric heating\\
   LW & 11.2 & 13.6 & $10^3$ & $\rm H_2$ dissociation\\
   EUV$_{\rm HI,1}$ & 13.6 & 15.2 & $10^3$ & HI ionisation\\
   EUV$_{\rm HI,2}$ & 15.2 & 24.59 & $10^3$ & HI and $\rm H_2$ ionisation\\
   EUV$_{\rm HeI}$ & 24.59 & 54.42 & $10^3$ & HeI ionisation\\
   EUV$_{\rm HeII}$ & 54.42 & $\infty$ & $10^3$ & HeII ionisation\\
        \hline
   \end{tabular}
   \label{tab:photons}
\end{table}

\subsection{Non-equilibrium photo-chemistry and radiative cooling}

The public version of {\sc ramses-rt} can solve non-equilibrium chemistry of hydrogen 
and helium species ($\rm HI$, $\rm HII$, $\rm HeI$, $\rm HeII$, $\rm HeIII$, and $\rm e^-$), 
involving collisional excitation, ionisation, and photoionisation \citep{rosdahl13}.
In order to take into account cooling by molecular hydrogen ($\rm H_2$), 
which is essential to model gas collapse in mini-haloes, 
we have made modifications based on \citet{glover10} and \citet{baczynski15}.
Note that the photon number density and fluxes in the eight energy bins 
that we describe in Table~\ref{tab:photons} are computed in a self-consistent way 
by tracing the photon fluxes from each star particle. The chemical reactions and radiative cooling 
are fully coupled with the eight photon groups. 
More details of the photo-chemistry will be presented in \citet{katz16} in terms of the prediction of 
molecular hydrogen in high-z galaxies.

Molecular hydrogen mainly forms on the surface of interstellar dust grains. 
However, in the early universe where there is little dust \citep[e.g.,][]{fisher14}, 
the formation is dominated by the reaction involving $H^-$. These hydrogen 
molecules are dissociated by  
Lyman-Werner photons with energy $11.2\,{\rm eV} \le E \le 13.6\,{\rm eV}$ 
or collisions with other species ($\rm HI$, $\rm H_2$, $\rm HeI$, and $e^-$).
Furthermore, $\rm H_2$, can be ionised by photons with $E\ge15.2\,{\rm eV}$.
We assume that all $\rm H_2^+$ is immediately destroyed by dissociative recombination.
To estimate the formation rate of molecular hydrogen through the $H^-$ channel,
we assume that the abundance of $H^-$ is set by the equilibrium between the 
formation and destruction via associative detachment and mutual neutralisation 
\citep[e.g.,][]{anninos97}, as
\begin{equation}
k_1 n_{\rm HI} n_{e} = k_2  n_{\rm H^-} n_{\rm HI}  + k_5  n_{\rm H^-} n_{\rm HII} + k_{13} n_{\rm H^-} n_{e},
\end{equation}
where $k_X$ is the reaction rate (reactions 1, 2, 5, and 13 in Appendix B of \citealt{glover10}).
This neglects the photo-detachment of $H^-$ by infrared photons, 
and it may thus lead to the over-estimation of the $H^-$ abundance when Pop II stars 
are present \citep{cen16}. However, we expect that cooling is dominated 
by metals once Pop III stars explode \citep[see Figure 2 of][for example]{wise14},
hence the gas collapse in mini-haloes is unlikely to be significantly affected.

\begin{figure*}
   \centering
   \includegraphics[width=14cm]{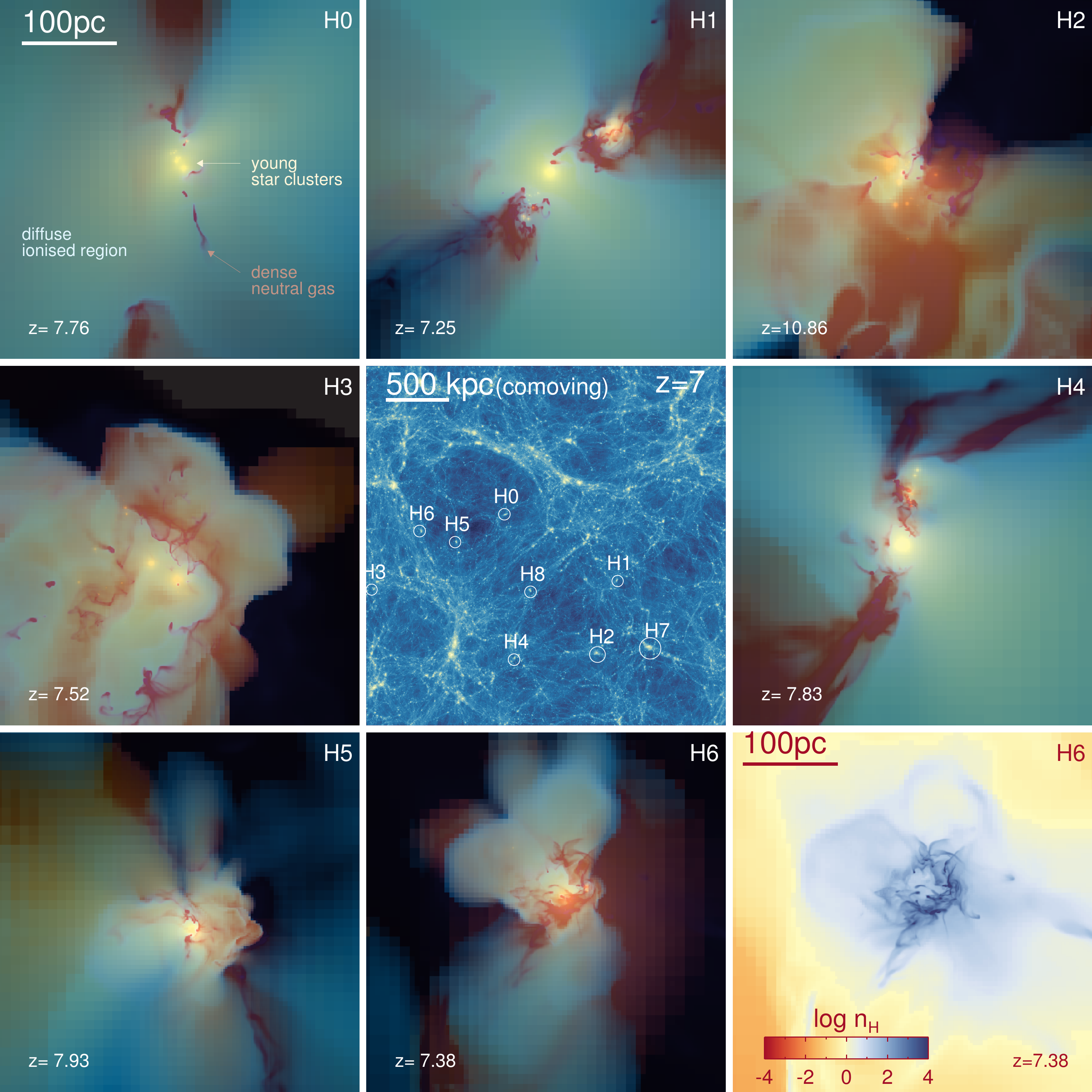} 
   \caption{
      Composite images of the density, HII fraction, and HI-ionising photon 
   density for seven simulated haloes out of the total nine sample. 
   The central panel displays the location of the nine zoom-in haloes in the entire 
   simulation box of length $2\,{\rm Mpc\,h^{-1}}$ (comoving) at $z=7$. 
   We evolve each halo until its mass becomes large enough ($\mhalo\approx10^8\,\msun$) 
   to radiate energy away mainly by atomic transitions. Actively star-forming regions 
   are shown as bright yellow colours, while highly ionised regions are displayed as 
   light blue colours. Dark regions show mostly neutral gas. It can be seen that the 
   central star-forming clump is disrupted by stellar feedback in many cases,
   leading to a high escape fraction of LyC photons.
   The bottom right most panel shows the projected gas density distribution of the 
   H6 halo for comparison.}
   \label{fig:pic}
\end{figure*}

Radiative cooling by hydrogen and helium species is directly computed from 
the chemical network. In particular, we include the cooling by molecular hydrogen 
following \citet{halle13}, which is largely based on \citet{hollenbach79}. 
In addition, Lyman Werner photons also heat the gas when they 
photo-dissociate and photo-ionise $\rm H_2$ or when they indirectly excite the 
vibrational levels of $\rm H_2$ \citep[see section 2.2.4 in][]{baczynski15}. 
Gas can cool further with the aid of metals,
which we consider by interpolating look-up tables which are pre-computed with the Cloudy 
code \citep{ferland98} as a function of density, temperature, and redshift \citep{smith11}.
Finally, we also include photoelectric heating on dust by UV photons with 
$\rm 5.6\,eV \le h\nu \le13.6\, eV$ \citep{bakes94} following \citet[][Section 2.2.5]{baczynski15}, as
\begin{equation}
\mathcal{H}_{\rm pe} = 1.3\times10^{-24} \,\epsilon G_0\,f_{\rm D/G} \, n_{\rm H} \,{\rm [erg\,cm^{-3}\,s^{-1}]},
\end{equation}
where $G_0$ is the strength of the local intensity in each cell, normalised to the 
Habing field \citep[$1.6\times10^{-3}\,{\rm erg\,s^{-1}\,cm^{-2}}$,][]{habing68},
and $f_{\rm D/G}=1$ is the dust-to-gas ratio, normalised to the local ISM value \citep{draine07b}. 
The efficiency for the heating ($\epsilon$) is taken from \citet{wolfire03}, as
\begin{align}
\epsilon = & \frac{4.9\times10^{-2}}{1+4.0\times10^{-3}\left(G_0 T^{1/2}/n_e \phi_{\rm PAH}\right)^{0.73}} \nonumber \\
 & \frac{3.7\times10^{-2} \left(T/10^4\right)^{0.7}}{1+2.0\times10^{-4} \left(G_0 T^{1/2} / n_e \phi_{\rm PAH}\right)},
\end{align}
where $\phi_{\rm PAH} = 0.5$ is a factor that controls the collision rates between electron and PAH. 
We do not use any uniform UV or Lyman Werner background radiation \citep[e.g.,][]{haardt12}.

\begin{figure}
   \centering
   \includegraphics[width=8.5cm]{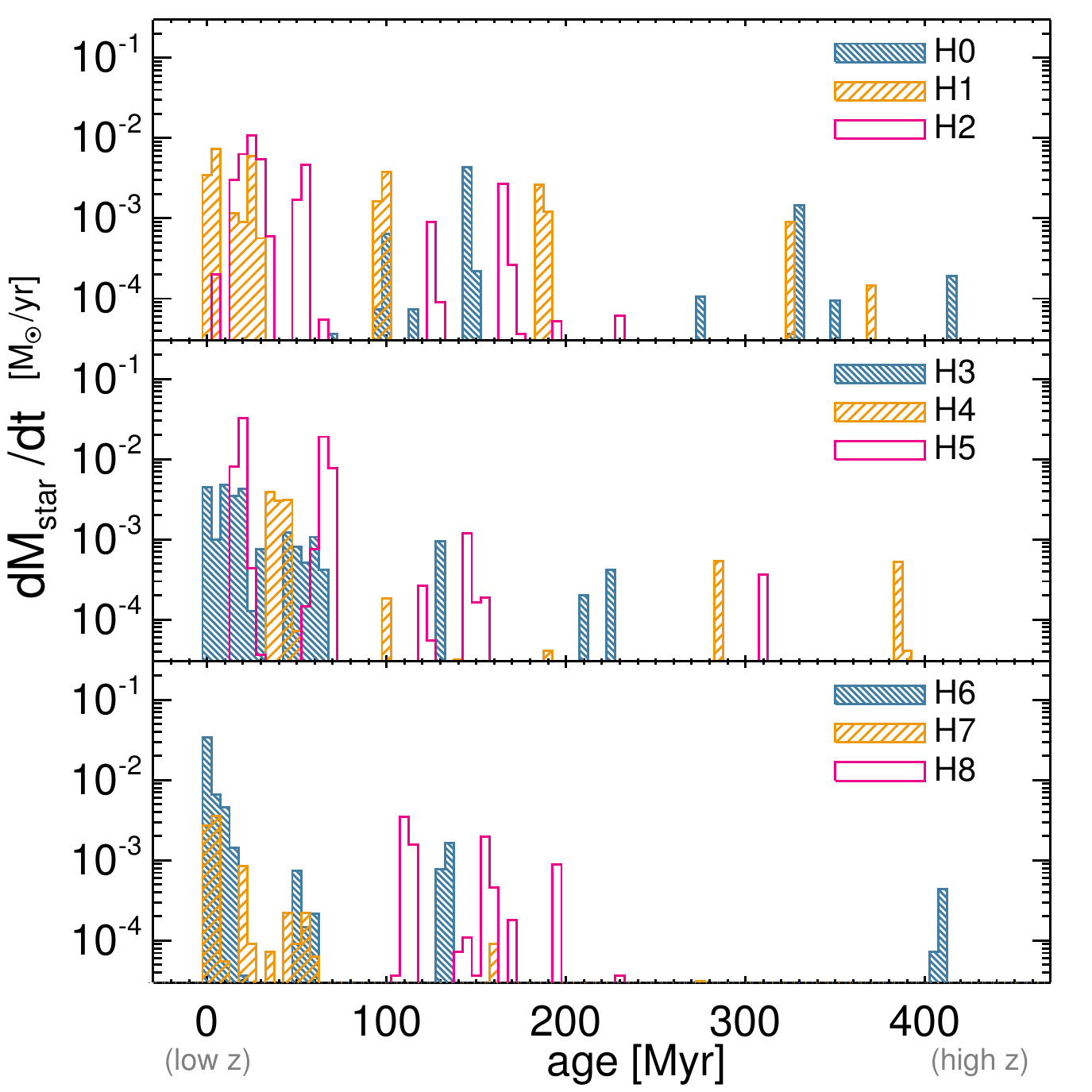} 
   \caption{ Star formation histories of the dwarf galaxies in our simulated mini-haloes.
   The x-axis indicates the age of stars measured at the end of each simulation. 
   Different colour coding denotes different galaxies. We split the sample for clarity.
   It can be seen that star formation is very stochastic. The recovery time 
   from the stellar feedback ranges from $\sim20 - 200$ Myr.}
   \label{fig:sfh}
\end{figure}

\section{Results}

The main aim of this paper is to assess the contribution  of mini-haloes to the 
reionisation history of the universe. For this purpose, we investigate the evolution 
of nine dwarf galaxies in haloes of mass $10^6\,\msun\la \mvir \la 10^8 \,\msun$ 
at $7\le z \le 20$. In this section, we first describe the main features of the simulated 
galaxies, present the escape fraction of LyC photons at the virial radius, 
and discuss the physical processes governing the evolution of the escape fraction.

\subsection{Galactic properties of the dwarf population during reionisation}

Our simulated haloes begin forming Pop III stars when the halo mass approaches 
a few times $10^6\,\msun$.  These Pop III stars disperse dense gas clouds 
and pollute the ISM and IGM with metals via energetic explosions. 
We find that the typical 
 metallicity of the halo gas after the explosion of Pop III stars is 
 $\sim 10^{-3} - 10^{-2} Z_\odot$, consistent with previous studies \citep{greif10,ritter12}. 
 The enrichment of the dense medium ($\nH\ge100\,\cmq$) takes place more slowly 
 than for the IGM  ($\sim 10^{-4} - 10^{-3} Z_\odot$), as this gas mixes with the newly accreted, 
 pristine material 
with primordial composition. Once the metal-enriched gas collapses, Pop II stars form 
in a very stochastic fashion. 
As the haloes become massive ($\sim10^8\,\msun$) and a large amount of gas 
accumulates in the halo centres, the star formation histories become less bursty, 
compared to those in haloes with masses of a few times $10^7 \,\msun$. 
Because stellar feedback violently disrupts star-forming clouds, the gas component of these 
mini-haloes show irregular morphologies rather than well-defined disks (Figure~\ref{fig:pic}). 
The resulting stellar metallicities in 
haloes with $\sim10^8\,\msun$ range from $Z=10^{-3}\,Z_\odot$ to $Z=10^{-1.7}\,Z_\odot$ with a median of $Z=10^{-2.6}\,Z_\odot$. 
We summarise several galactic properties in the nine simulated haloes in Table~\ref{tab:sim_summary}.

\begin{table*}
   \caption{Summary of simulation results. All quantities are measured at the final redshift of each  simulation.
   Column (1): ID of simulated halo.
   Column (2): virial mass of the dark matter halo.
   Column (3): total stellar mass of Pop II.  
   Column (4): total gas mass inside a halo. 
   Column (5): mass-weighted mean metallicity of star particles.
   Column (6): mass-weighted mean gas metallicity inside a halo.
   Column (7): total stellar mass of Pop III. 
   Column (8): initial redshift to form Pop III stars.
   Column (9): initial redshift to form Pop II stars.
   Column (10): final redshift of each simulation. 
   }
   \centering
   \begin{tabular}{cccccccccc}
   \hline
  Halo ID & $\log \mhalo$ & $\log M_{\rm star, II}$ & $\log M_{\rm gas}$ & $\left<\log Z_{\rm star}\right>$ & $\log Z_{\rm gas}$ &  $\log M_{\rm star, III}$ &  $z_{\rm PopIII}$ & $z_{\rm PopII}$ & $z_{\rm final}$ \\
               & [$\msun$] &  [$\msun$]  &  [$\msun$]   & [$\Zsun$]  &  [$\Zsun$]  &  [$\msun$] & & \\
     \hline
    H0 & 8.05 & 4.53 &  6.93  &  -3.0  &  -2.7  & 3.20 &  12.4  & 10.6     & 7.0\\
    H1 & 8.03 & 5.17 &  7.10  &  -2.8  &  -2.4  & 1.57 &  14.3  & 12.7     & 7.3\\
    H2 & 8.08 & 5.26 &  7.06  &  -2.5  &  -2.0  & 3.35 &  18.1  & 15.0   & 10.6\\
    H3 & 8.00 & 5.08 &  7.16  &  -2.7  & -2.5   & 1.40 &  10.9   & 10.0   & 7.5\\
    H4 & 8.03 & 4.72 &  6.32  &  -2.8  & -2.6  & 3.50 & 14.7    & 13.4  & 7.0\\
    H5 & 8.01 & 5.17 &  6.98  &  -2.7  & -2.3  & 3.31 &  14.7    & 13.6  & 8.1\\
    H6 & 7.96 & 5.70 &  7.17  &  -2.4  & -2.0  & 2.16 &  15.6   & 13.8  & 7.3\\
    H7 & 7.91 & 4.61 &  7.09  &  -3.0  & -2.9  & 2.69 &  13.4   & 12.4  & 11.2\\
    H8 & 7.85 & 4.17 & 7.00   & -3.3   & -3.2  & 1.59 & 12.5 & 10.7   & 8.1 \\
       \hline
   \end{tabular}
   \label{tab:sim_summary}
\end{table*}

\begin{figure}
   \centering
   \includegraphics[width=8.5cm]{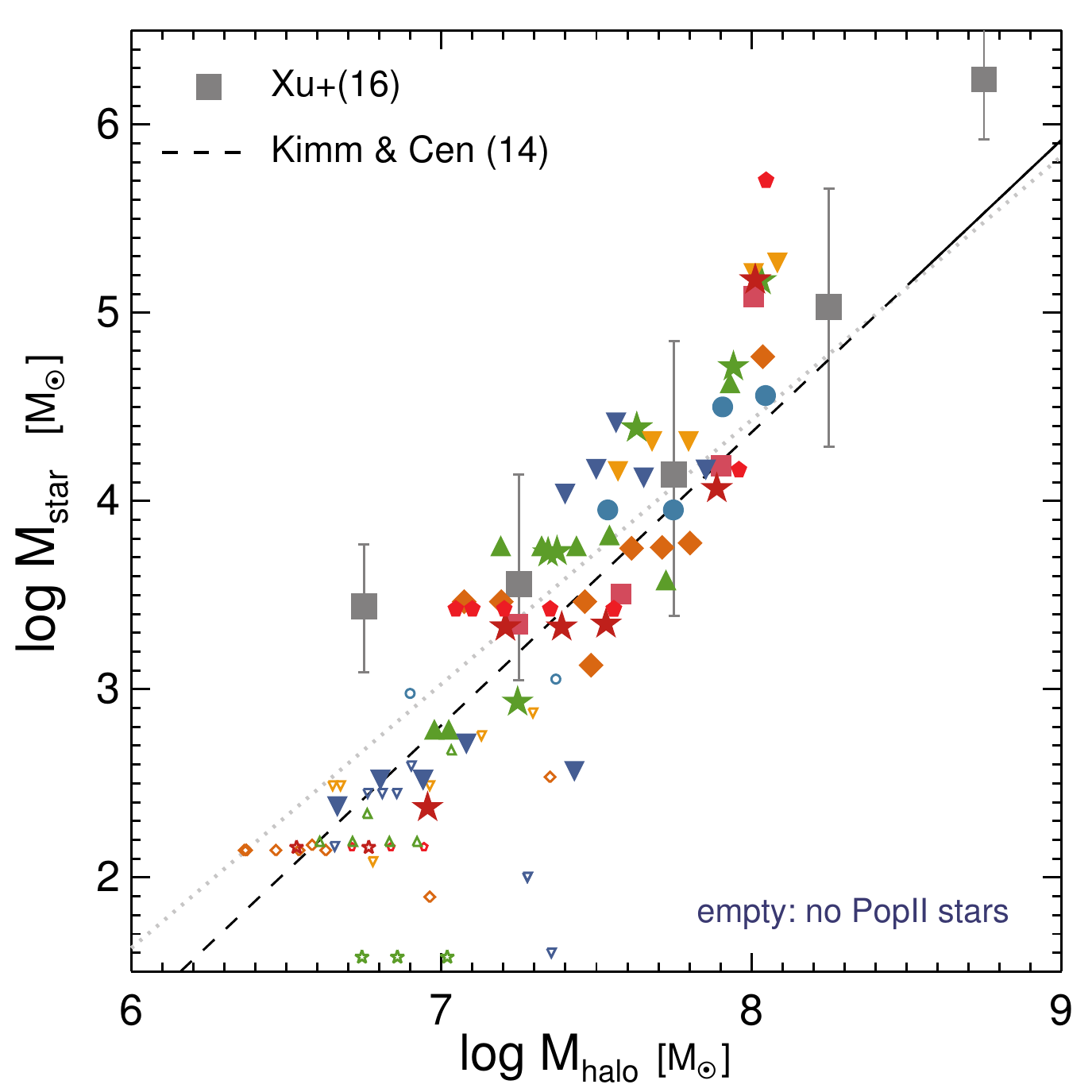} 
   \caption{
   Predicted stellar mass as a function of halo mass in our simulations. 
   Different colour codings and symbols correspond to nine different haloes. Note that we plot 
   the results at various redshifts, and thus this may also be seen as an evolutionary 
   sequence at $7\le z< 18$. The empty symbols indicate the haloes 
   hosting only Pop III stars, while the haloes with Pop II stars are shown as filled symbols.
   We also include the stellar mass in haloes outside the main halo 
   if they are still within the zoom-in region and not contaminated by coarse dark matter 
   particles.
   We find that our results follow the slope and normalisation predicted by 
   \citet{kimm14} (black line) which reproduced the UV luminosity 
   function at $z\sim7$. The dashed line indicates an extrapolation of the \citet{kimm14} results.
   Our results are also in fair agreement with simulations from \citet{xu16} (the grey squares).
   For comparison, we include the empirical 
   sequence at $z\approx0$ extrapolated to the mini-halo regime \citep{behroozi13} 
   (grey dotted line).
    }
   \label{fig:mstar}
\end{figure}

\begin{figure}
   \centering
   \includegraphics[width=8.5cm]{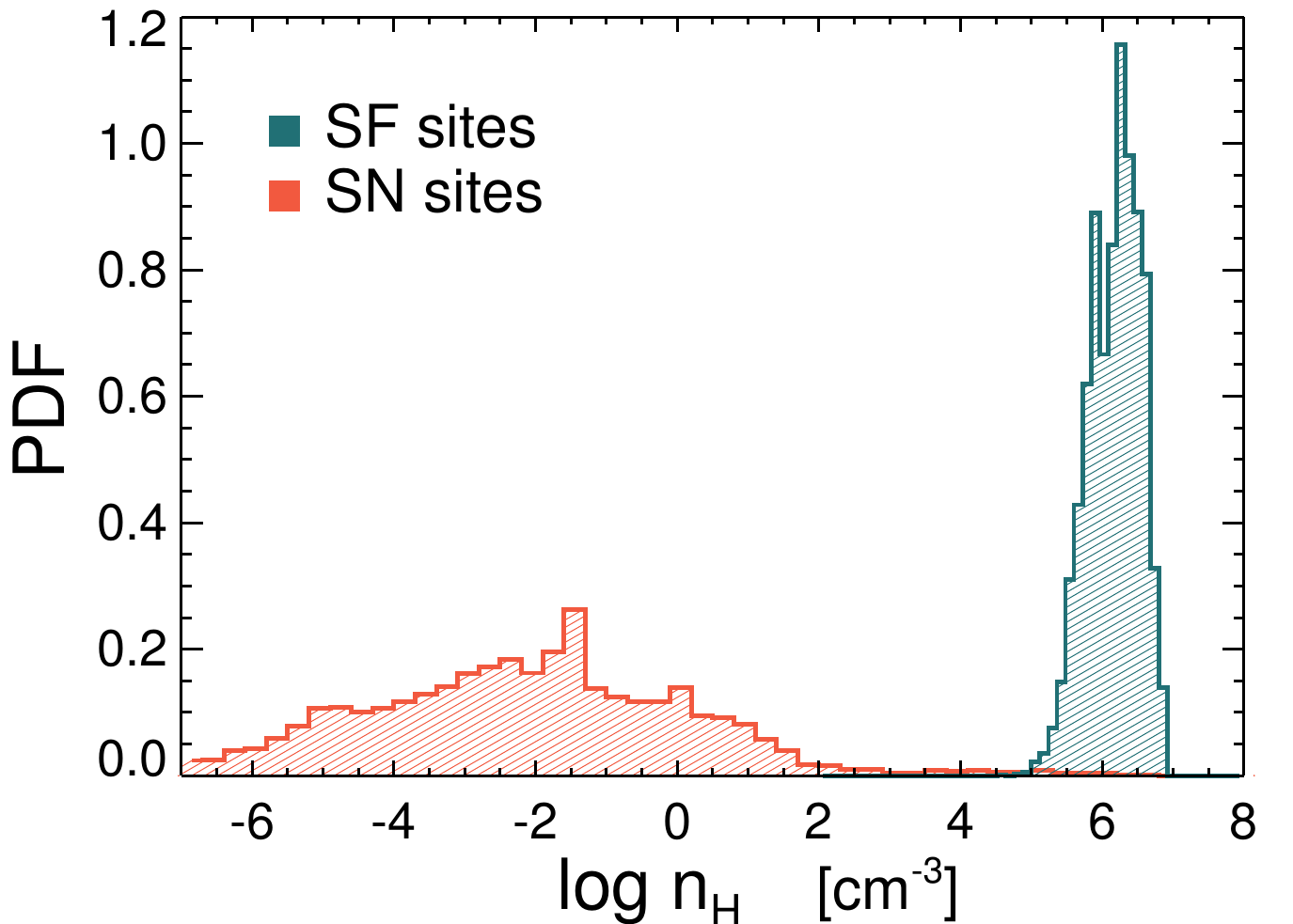} 
   \caption{Local environments of star formation and Type II SN explosions 
   in our mini-halo simulations. Even though stars form in very dense media, 
   SN explosions occur in low-density regions. This is firstly because radiation lowers the 
   ambient density before SN explosions, and secondly because the lifetime of 
   massive stars ranges from 3 to 40 Myr and late explosions take advantage 
   of the early SN events.}
   \label{fig:sites}
\end{figure}

Overall, we find that {\em  star formation is very inefficient in the mini-haloes} (Figure~\ref{fig:mstar}). 
For example,  haloes with masses $\sim10^{7.5}\,\msun$ form clusters of stars of 
$\sim10^{3-4}\,\msun$.  At larger masses ($\mvir\sim10^8\,\msun$), the efficiency becomes 
higher, but still $\la 1\%$ of baryons are converted into stars. We note that these 
results are consistent with the recent radiation-hydrodynamic calculations of \citet{wise14,xu16}.
However, our results are slightly different from these studies in the sense 
that the least massive haloes ($\mvir\la10^7\,\msun$) appear to host progressively 
smaller amount of stars, whereas, in \citet{wise14}, the stellar mass appears to 
saturate at a few times $10^{3}\msun$. 
We also find that the dispersion in the stellar mass-halo mass relation is  
smaller than that of \citet{xu16}. This is likely to be due to the small number of
 samples used in this work. Indeed, \citet{xu16} find a larger 
dispersion than \citet{wise14} when they increase the number of galaxies 
from 32 to $\sim$ 2000 simulated with the same assumptions.

The inefficient star formation in our simulations is due to strong stellar feedback. 
This can be inferred from Figure~\ref{fig:sfh} where the star formation histories are 
shown to be very bursty. The typical time scale of star formation does not exceed 
$\sim$ 10 Myr and is often smaller than 5 Myr  when the halo mass is small. 
The recovery time after a 
burst of star formation in the mini-haloes is also large, ranging from 
$\sim$ 20 Myr to $\sim$ 200 Myr. 
Because radiation can drive winds by over-pressurising the ISM and
early SNe can create cavities in the star-forming regions, 
we find that the majority of SNe explode in 
a low-density environment  with  $\nH \sim 2\times10^{-3}\,{\rm \cmq}$, while stars form only 
in very dense environments with $\nH \ge 4\times10^4\,\cmq$ (Figure~\ref{fig:sites}).  

At present, there are no direct observational constraints on the evolution of galaxies 
in mini-haloes at high redshift, and thus it is not currently possible to validate our feedback model. 
Nevertheless, it is worth noting that the simulated galaxies follow the stellar mass-to-halo 
mass sequence obtained from \citet{kimm14}  (Figure~\ref{fig:mstar}), which successfully 
reproduces the faint-end slope and normalisation of the observed UV luminosity function at $z\sim7$ 
without dust correction.
It is also interesting to note that the predicted stellar masses are comparable 
to the {\it local} stellar mass-to-halo mass relation derived from the abundance matching 
technique \citep{behroozi13} when extrapolated to the mini-halo mass regime. 
The fact that the simulated galaxies are very metal-poor ($\sim 0.003 \,Z_\odot$) 
suggests that our results do not suffer from the over-cooling problem,
as it usually leads to significantly higher metallicities of $\ga 0.1 \,Z_\odot$ \citep{wise12b}.

Our simulated galaxies are slightly more metal-poor (roughly a factor of two) than 
the local dwarf population \citep{woo08}. However, given the large scatter in the observed 
stellar metallicities, the difference is unlikely to be significant.
Rather, we find that the simulated protogalaxies are more compact (by a factor of a few) than 
 the local dwarf spheroids of comparable masses \citep[e.g.,][]{brodie11}. 
As can be observed in Figure~\ref{fig:pic}, the stellar components of high-z dwarves are 
dominated by a few clusters, and the resulting half-mass radii are found to be $\sim$ 20-100 pc.
The smaller size of the simulated galaxies may not be very surprising, given that the universe is denser 
and star formation per unit stellar mass is known to be more efficient at high redshifts \citep[e.g.,][]{speagle14}.

\subsection{Escape Fraction of LyC photons}

We calculate the escape fraction by comparing the photon flux ($F_{\rm ion}$) 
at the virial radius with the emergent flux from stars. Since photons travel with finite speed, 
we use the production rate at $t- \rvir / \tilde{c}$, where the time delay ($\rvir / \tilde{c}$) is 
roughly $\sim 1$ Myr. The escape fraction is then
\begin{equation}
f_{\rm esc} (t)= \frac{\int \, d \Omega \, \vec{F}_{\rm ion} (t)\cdot \hat{r} }{\int dm_\star \dot{N}_{\rm ion} (t- \rvir/\tilde{c})},
\end{equation}
where $\Omega$ is the solid angle, $m_\star$ is the mass of star particles, 
$\dot{N}_{\rm ion}$ is the production rate of ionising radiation per unit mass.
Note that the choice of the radius (i.e., \rvir) at which the measurement is made 
is conventional, but it is desirable to measure the escape fraction at a radius 
large enough that it can be combined with estimates of the clumping factor 
from large-scale simulations
\citep[e.g.,][]{pawlik09,finlator12,shull12,so14} to study
the reionisation history of the universe.

\begin{figure}
   \centering
   \includegraphics[width=8.5cm]{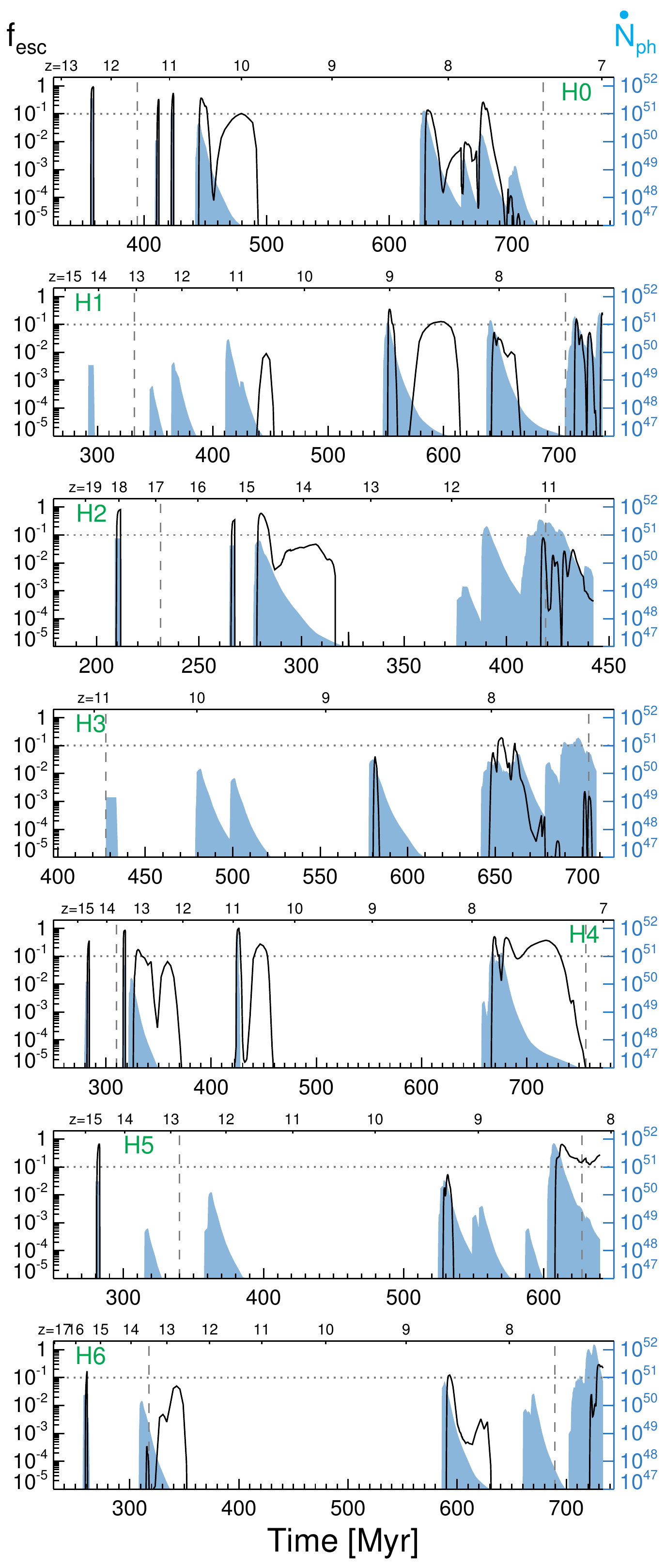} 
   \caption{ 
   Evolution of the escape fraction (black lines) and the photon production rates ($\dot{N}_{\rm ph}$)
   in units of $\#\,s^{-1}$ (cyan shaded regions) in seven different haloes as a function 
   of the age of the Universe. The two vertical dashed lines mark the time at 
   which the mass of each halo becomes $10^7$ or $10^8\,\msun$, respectively.
   The photons produced by Pop III and Pop II stars can be distinguished by the shape 
   of the photon production rate; only Pop III stars exhibit a squarish evolution, because 
   no LyC photons are generated once they explode or implode. 
   Mini-haloes show a high escape fraction in general, although a significant 
   variation can be seen. The escape fraction often exhibits a double peak 
   for individual star formation events, which is explained in detail later in Figure~\ref{fig:twopeak}.  
}
   \label{fig:fesc}
\end{figure}

\begin{figure}
   \centering
   \includegraphics[width=8.5cm]{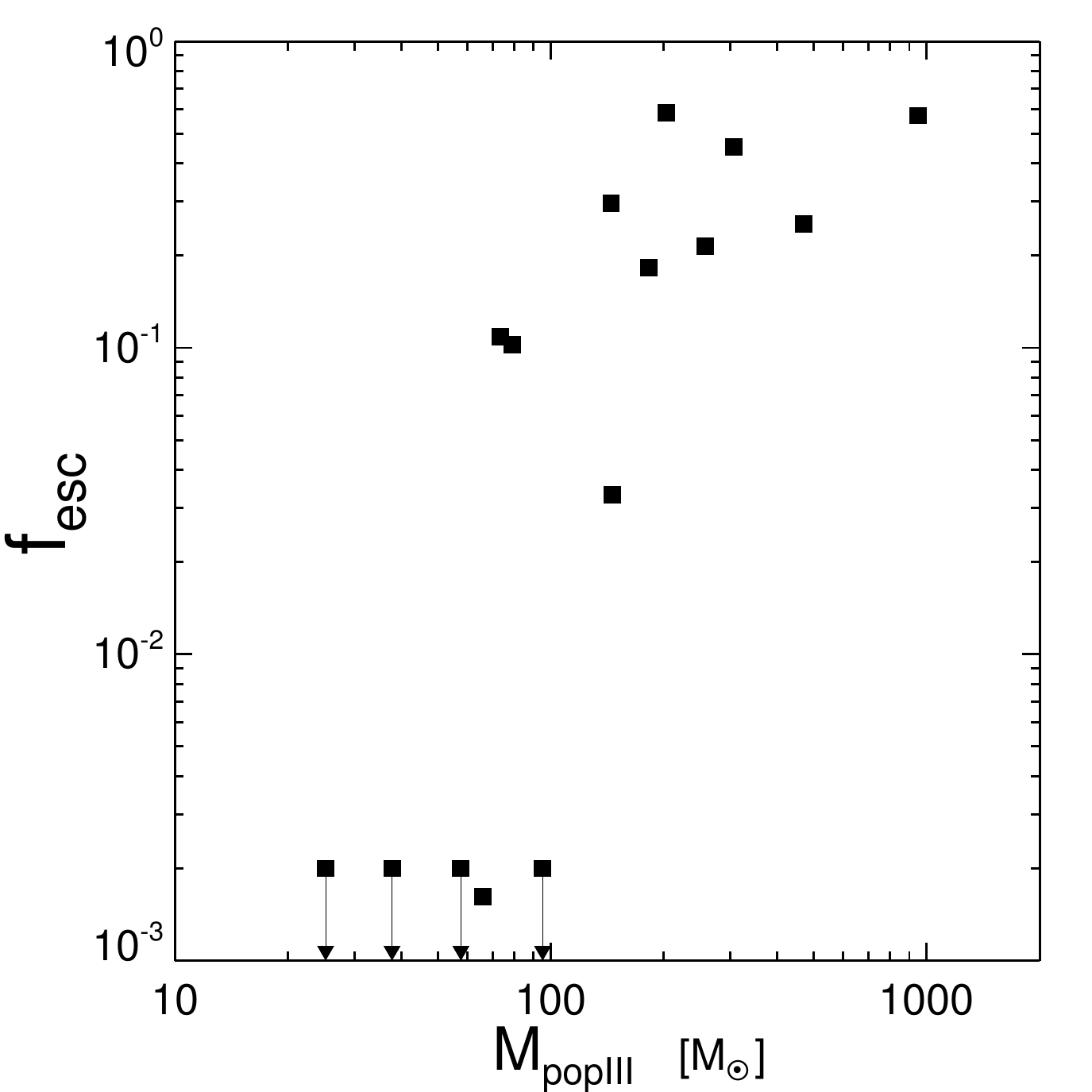} 
   \caption{Escape fraction of individual Population III stars. We measure the 
   mean escape fraction of Pop III stars by weighting the photon production rate 
   during their lifetime. More massive Pop III stars show a higher escape fraction. 
   We find that most LyC photons from low-mass Pop III stars ($M\la 60 \msun$) 
   are absorbed within the dark matter halo. }
   \label{fig:fesc_pop3}
\end{figure}

\subsubsection{A High Average Escape Fraction in  mini-haloes}

Figure~\ref{fig:fesc} shows that LyC photons escape from mini-haloes quite efficiently 
after a burst of star formation. Not only Pop III stars, which are characterised by a 
constant photon production rate and an abrupt decrease, but also Pop II stars, 
characterised instead by  an exponentially decaying rate, show a high escape fraction 
of $\fesc\sim30-40\%$.
In some cases, the escape fraction remains very low even after the formation of 
Pop II star clusters. Similarly, not all Pop III stars lead to a high escape fraction. 
Figure~\ref{fig:fesc_pop3} displays the photon-number weighted escape fraction 
of individual  Pop III stars with different masses. It can be seen that only massive Pop III 
stars with $M_{\rm Pop III} \ga 100 \,\msun$ are able to provide LyC photons 
to the IGM \citep{whalen04}, while almost all of the ionising radiation from Pop III stars 
with $M_{\rm Pop III} \la 70 \,\msun$ is absorbed 
inside the virial radius.

\begin{figure}
   \centering
   \includegraphics[width=8.2cm]{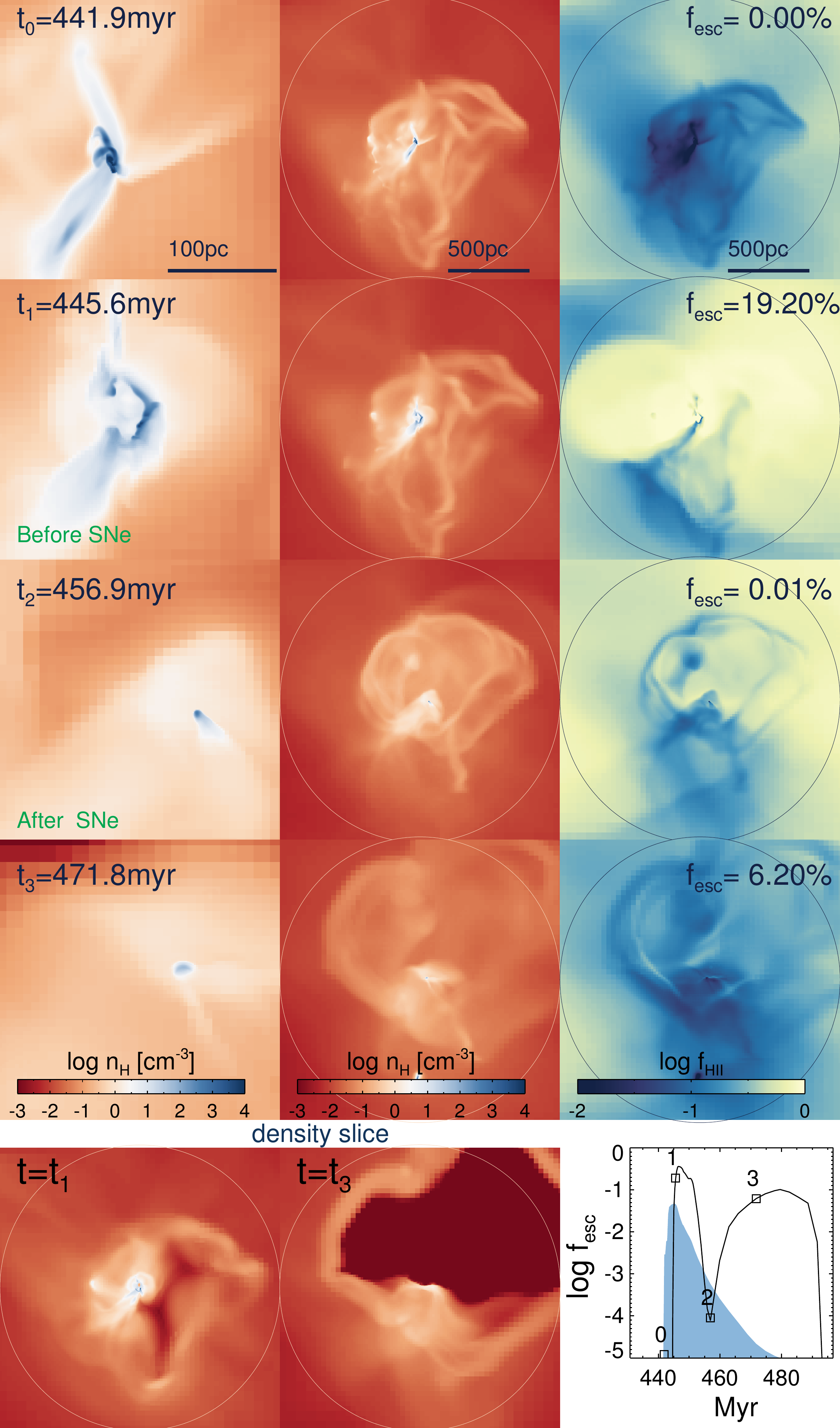} 
   \caption{Origin of the double-peaked escape fraction for individual star formation events. 
   The double-peaked feature is shown in the bottom right panel for clarity. 
   The first column of the mosaic shows the evolution of the
   projected density distributions in the central region of the halo at four different times, 
   as marked in the bottom right panel. The second and the third columns display 
   the projected density and the fraction of  ionised hydrogen distributions within 
   a virial radius. The escape fraction at each time is also indicated in the top right corner.
   The bottom left panels show the density slice at two different epochs.
   One can see that the escape fraction becomes very high once the central 
   star-forming clump is destroyed, and then drops as massive stars end their life.
   The escape fraction increases again when SNe blow out gas from the halo.   }
   \label{fig:twopeak}
\end{figure}

The high escape fraction can be associated with the blow out of birth clouds due to 
radiation feedback (i.e. photoionisation plus direct radiation pressure). 
This is especially evident for Pop III stars, as they tend to form in an isolated fashion and 
radiation is the only energy source while they are emitting LyC photons.
Even for Pop II star clusters, we find that radiation feedback is the main culprit 
for creating the low-density, ionised channels through which LyC photons can escape.  
This is supported by the short time delay ($\la 5$ Myr) between the peak of the 
photon production rate and the peak of the escape fraction.  Even though the youngest 
SN occurs after  3.5 Myr, the stochasticity in our random sampling of the lifetime of SN 
progenitors is unlikely to explain the short delay. To substantiate this further, we show 
an example in Figure~\ref{fig:twopeak} where the escape fraction increases 
from $\fesc(t_0)=0\%$ to $\fesc(t_1)\sim20\%$ within 3.7 Myr, during which no 
SN explosions occur. The dense, star-forming clouds are disrupted and LyC photons 
propagate to the virial radius, ionising the neutral hydrogen in the halo (second row). 
Note that only the birth clouds are dispersed, while the average density 
of the halo gas is little affected by radiation.

We find that SN explosions enhance the escape of LyC photons 
from time to time by ejecting gas from the dark matter halo. As an illustration, 
in Figure~\ref{fig:twopeak}, we show the projected density distributions and the ionisation fraction 
of hydrogen at several different epochs. 
After the birth clouds are dispersed and lifted by radiation feedback ($t=t_1$),
the density of the gas beyond the galaxy actually increases, obscuring the LyC photons 
in the halo region ($t=t_2$). Once this gas is completely pushed out from the halo, 
the column density of neutral hydrogen along these solid angles becomes very small 
($t=t_3$, bottom middle panel). As a result, even though the projected ionised fraction 
at $t=t_3$ appears to be smaller than at the previous stage ($t=t_2$), the actual escape 
fraction is larger. Nevertheless, the effect of SNe by creating the secondary peak 
does not play a significant role in increasing the total number of escaping photons,
as the stellar populations become too old to generate a large amount of LyC photons.
It should be noted, however, that the effects of SNe may be more substantial 
if several star clusters form  simultaneously in more massive haloes 
($\mhalo \ga 10^8\,\msun$) and SNe in slightly older clusters generate 
strong winds that  strip off gas in other star-forming clumps \citep[e.g.,][]{kimm14}.


\begin{figure}
   \centering
   \includegraphics[width=8.5cm]{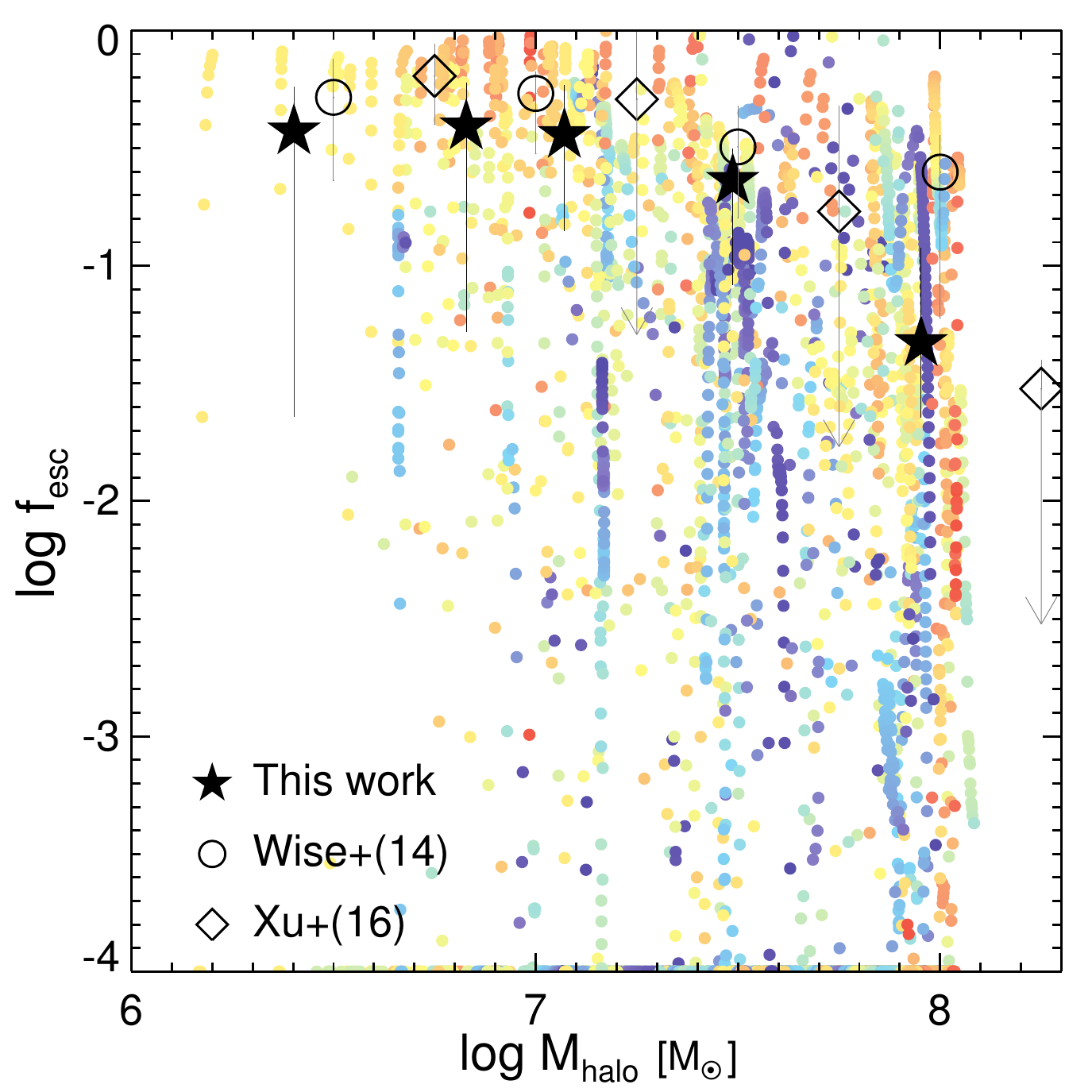} 
   \caption{ Escape fraction of LyC photons as a function of halo mass. 
   Each point corresponds to the escape fraction in different haloes at various redshifts.
   Larger photon production rates are shown as redder colours.
   The photon number-weighted average is displayed as the star symbols 
   with error bars representing the interquartile range (see Table~\ref{tab:fesc}). 
   We find that the photon-weighted escape fraction is high ($\sim 20$--$40\%$)
   in the mini-haloes, in good agreement with the previous studies \citep{wise14,xu16},
   even though there is a considerable scatter in instantaneous measurements.
   }
   \label{fig:fescavg}
\end{figure}

In Figure~\ref{fig:fescavg}, we show the average escape fraction as a function of the
dark matter halo mass. As demonstrated in previous studies 
\citep{kimm14,wise14,paardekooper15}, the instantaneous escape fraction varies 
significantly for a given halo mass. However, the photon-number weighted mean 
escape fraction in the mini-halo regime ($\mvir\le 5\times10^7$) 
is found to be generally very high ($\sim 20$--$40\%$, Table~\ref{tab:fesc}). Note that this is a factor of 
a few larger than the average escape fraction predicted in atomic-cooling haloes 
\citep[$\fesc\sim10\%$,][]{kimm14}, indicating that there is a dependence on the halo mass.
It is also interesting to note that our results are in good agreement with \citet{wise14,xu16},
despite the significant differences in the modelling of star formation and feedback. 
Nevertheless, both sets of studies (this work and \citet{wise14,xu16}) allow for rapid star 
formation which leads to the disruption of gas clumps. \citet{wise14} adopted 7\% efficiency 
for star formation within a sphere of mean density $\nH\sim 10^3\, \cmq$ per dynamical time, 
while our simulations employ an efficiency that varies ($\sim$ 5 -- 20\%) according to the local turbulent 
and gravitational conditions. The common prediction of the high $\fesc$ suggests that 
there is a dominant feedback process, which is included in both studies, shaping 
the escape fraction in the mini-haloes.
Indeed, as will be detailed later, we find that photoionisation is the main culprit
for the effective leakage of LyC photons (see Section 4).

\begin{table} 
\caption{Photon number-weighted $f_{\rm esc}$ at $7\le z \la 15$}
\centering
\begin{tabular}{@{}cc}
\hline 
$\log M_{\rm vir}$ & $\left<f_{\rm esc}\right>$       \\
\hline 
6.40     & 0.373$^{+0.206}_{-0.350}$\\
6.83     & 0.391$^{+0.211}_{-0.339}$ \\
7.07    & 0.360$^{+0.228}_{-0.219}$\\
7.49    & 0.230$^{+0.086}_{-0.074}$\\
7.95    & 0.050$^{+0.147}_{-0.026}$\\
\hline 
\label{tab:fesc}
\end{tabular} 
\end{table}

\subsubsection{Low Escape Fraction Due to Slow Destruction of Birth Clouds}

\begin{figure}
   \centering
   \includegraphics[width=8.5cm]{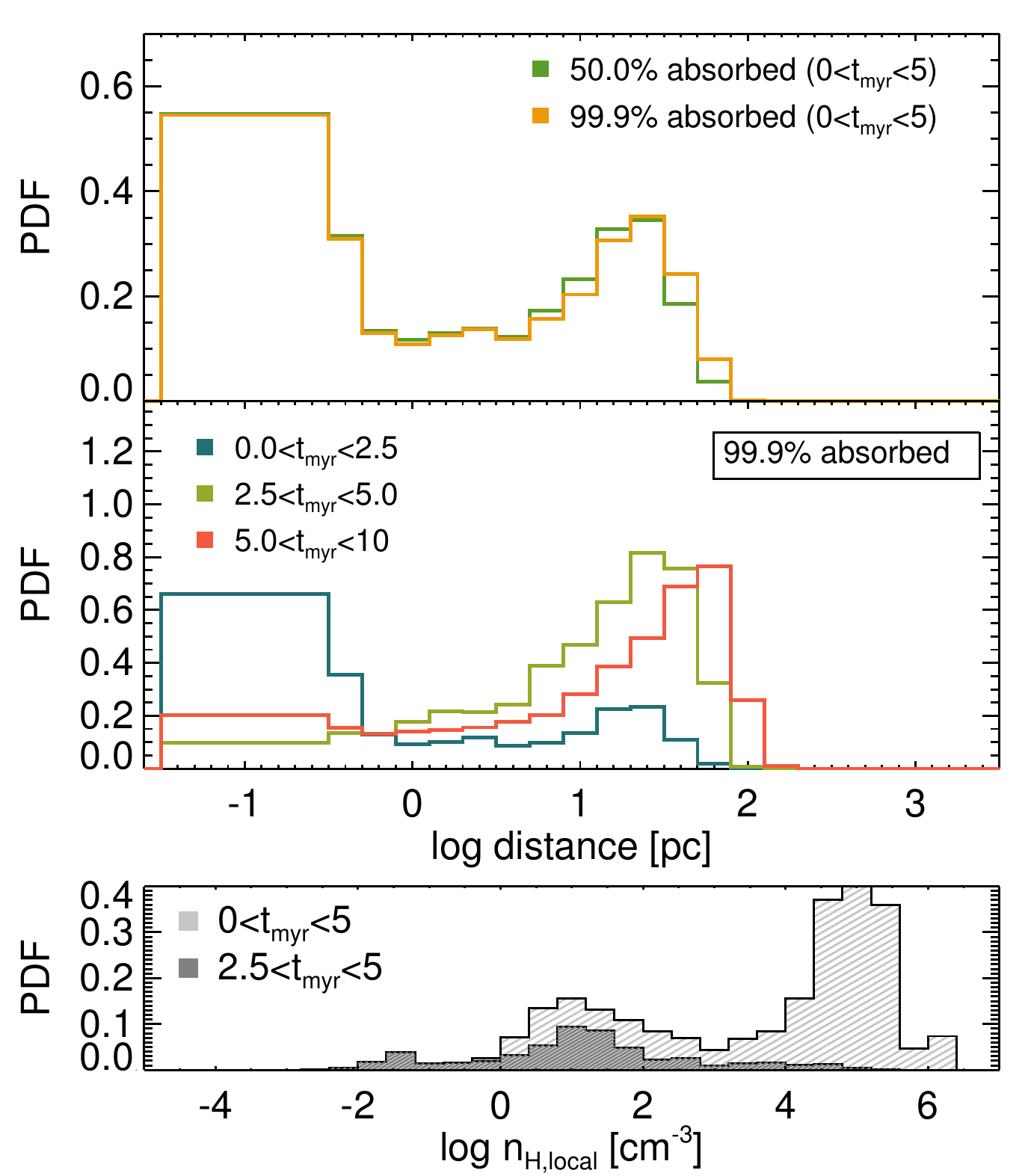} 
   \caption{{\it Top panel}: probability distribution functions (PDFs) of the distance 
   within which photons from stars younger than 10 Myr are absorbed
   when the escape fraction is low ($\fesc<10^{-4}$) even after starbursts. 
   These, for example, correspond to $11 < z <11.5$ in H2 or $7 < z < 8$ in H6 (see text).  
   Also included as a dark red line is the PDF of the distance within which 99.9\% of the 
   photons from stars of age $5< t <10$ Myr is absorbed.  
   {\it Bottom panel}: probability distribution function of the local density 
   at which the young stars are located. }
   \label{fig:pdf_low}
\end{figure}

Even though the leakage of LyC photons in the mini-haloes is very efficient on average, 
mini-haloes are sometimes optically thick for the LyC photons to escape (Figure~\ref{fig:fesc}). 
In order to understand the origin 
of the occasionally low escape fraction, we compute the optical depth centred on each young ($<10$ Myr) 
star by casting 3072 rays per stellar particle using the HealPix algorithm \citep{gorski05}. 
We adopt the absorption cross-section for neutral hydrogen from \citet{osterbrock06}, and the 
Small Magellanic Cloud-type extinction curve for dust \citep{weingartner01} assuming 
that the dust mass constitutes 40\% of the metal mass \citep[e.g.,][]{dwek98,draine07}. 
We neglect the contribution from metals residing in hot gas with $T>10^6\,{\rm K}$, 
as they are likely to be thermally sputtered \citep[e.g.,][]{draine79}.

In Figure~\ref{fig:pdf_low}, we examine the minimum distance 
within which 50\% and 99.9\% of the LyC photons would be absorbed. 
Also included in the bottom panel is the local density at which 
each young star is located. Since we are interested in the origin of the low escape fraction  
even after a burst of star formation, we restrict our analysis to $11 < z < 13$ in H1, 
$11.2 < z < 12$ in H2, $9.5 < z < 10$, $7<z<7.5$ in H3, and
$7.5<z<8$ in H6. The plot demonstrates that the escape fraction is low in two circumstances.
First, a  high density of the local environment ($\nH\ga10^3\,\msun$) indicates that 
stars are either just born out of dense clouds or radiation is not strong enough to blow 
away the surrounding medium. In this case the rays from these young star particles typically 
propagate to a small distance ($r\la 10$ pc), implying that they get 
absorbed within the birth clouds. Such stars account for $\sim70\%$ of the low escape fraction.
Second, stars with older ages ($t \ga 2.5\,{\rm Myr}$) show progressively 
larger minimum distances (the middle panel), meaning that the birth clouds are dispersed 
at later stages. The trend that older stars are located in lower density environments
supports this interpretation (the bottom panel). Once the dense structures are destroyed, 
photons propagate out to a larger radius ($r\ga10\,{\rm pc}$), but only a small fraction 
of the total ionising radiation appears to be able to do so. The similarity of the density  PDFs for 
50~\% and 99.9~\% absorption probability  in the top panel suggests  that the photon luminosity 
is sufficiently weak for  the ionisation front to stall inside the halo. 
The composite image of H5 in Figure~\ref{fig:pic} illustrates an example of such a case.
Thus, the escape fraction can sometimes be low
even after a burst of star formation mainly because the birth clouds 
are cleared away too slowly compared to the lifetime of massive stars ($\sim5 {\rm Myr}$). 

\section{Discussion}
\subsection{Physical mechanism for driving the escape of LyC photons in  mini-haloes}

In the previous section, we showed that the high escape fraction is best explained by strong 
radiation feedback. This raises  the question, which mechanism within this category is primarily  responsible for regulating 
the escape fraction (i.e., photo-heating, direction radiation pressure from UV photons, 
or multiply scattered IR photons)?
Whether or not multiply scattered IR photons 
govern the dynamics of the ISM in our primordial galaxies can be tested 
by examining the trapping factor in the dense regions of the ISM. 
The typical density of the star-forming clouds of size $\sim 10\,{\rm pc}$ 
is $\sim10^5\,\cmq$ in our simulations, and their metallicity ranges 
from $Z\sim10^{-4}\,Z_{\odot}$ to $10^{-2}\,Z_{\odot}$. Assuming a dust 
opacity of $\kappa_{\rm sc} \sim 5\,(Z/Z_\odot)\,{\rm cm^2\,g^{-1}}$ \citep{semenov03}, 
IR photons in the star-forming clouds have a maximum optical depth 
of $\tau_{\rm d} = \kappa_{\rm sc}\, \Sigma_{\rm gas} \la 0.3$.
Note that $\tau_{\rm d}$ will be reduced further if the smaller 
scattering cross-section is considered at 
temperatures lower than $\sim {\rm 100\,K}$ \citep{semenov03}.
This indicates that the IR photons are not efficiently trapped inside the ISM,
and that reprocessing gives a negligible momentum boost 
in the metal-poor, mini-haloes \citep[c.f., the min$\rho_Q$ case in ][]{bieri16}.

The effects of photo-heating and direct radiation pressure are difficult to disentangle,
because they operate simultaneously. Nevertheless, it is possible to estimate 
the maximum extent within which each process can balance the external pressure in the ISM.
\citet{rosdahl15a} show that if the ambient gas temperature ($T_{\rm 0}$) is significantly lower 
than that of the HII bubble ($T_{\rm ion}\approx 2\times10^4\,{\rm K}$), 
the gas density inside the bubble can be lowered due to over-pressurization, 
and the extent to which ionising radiation can balance recombination becomes larger 
by a factor of $(T_{\rm ion}/T_{\rm 0})^{2/3}$. Thus, the gas can be pressure-supported 
within $r_{\rm PH}$, 
\begin{align}
r_{\rm PH}  \approx 26\,{\rm pc} & \left( \frac{m_{\rm star}}{10^3 \msun} \right)^{1/3} \left( \frac{n_{\rm H,0}}{10^3\,{\rm cm^{-3}}} \right)^{-2/3} \nonumber \\
    & \left( \frac{T_{\rm ion}}{10^4\,{\rm K}} \right)^{2/3} \left( \frac{T_{\rm 0}}{10^2\,{\rm K}} \right)^{-2/3},
    \label{eq:rPH}
\end{align}
where $n_{\rm H,0}$ is the density of the ambient medium.
On the other hand, the absorption of ionising radiation directly imparts momentum 
which offsets the external pressure within $r_{\rm DP}$,
\begin{align}
r_{\rm DP} & = \sqrt{\frac{L}{4 \pi \,n_{\rm H,0} \,c \,k_{\rm B} \,T_0}} \nonumber \\
 & \approx 6.4\,{\rm pc} \left(\frac{m_{\rm star}}{10^3\,\msun}\right)^{1/2} \left(\frac{n_{\rm H,0}}{10^3\,{\rm cm^{-3}}}\right)^{-1/2} \left(\frac{T_0}{10^2\,{\rm K}}\right)^{-1/2}.
     \label{eq:rDP}
\end{align}
Here we use $L=2\times10^{36}\,{\rm erg\,s^{-1}}$ per 1 \msun, 
adequate for an SSP with age younger than 5 Myr.
As pointed out \citet{rosdahl15a}, the direct radiation pressure from a massive star cluster dominates 
over the pressure from the warm ionised gas only in a very dense medium,
as $r_{\rm PH} / r_{\rm DP} \propto \left( L\,n_{\rm H,0}\right)^{-1/6}$.

\begin{figure}
   \centering
   \includegraphics[width=8.5cm]{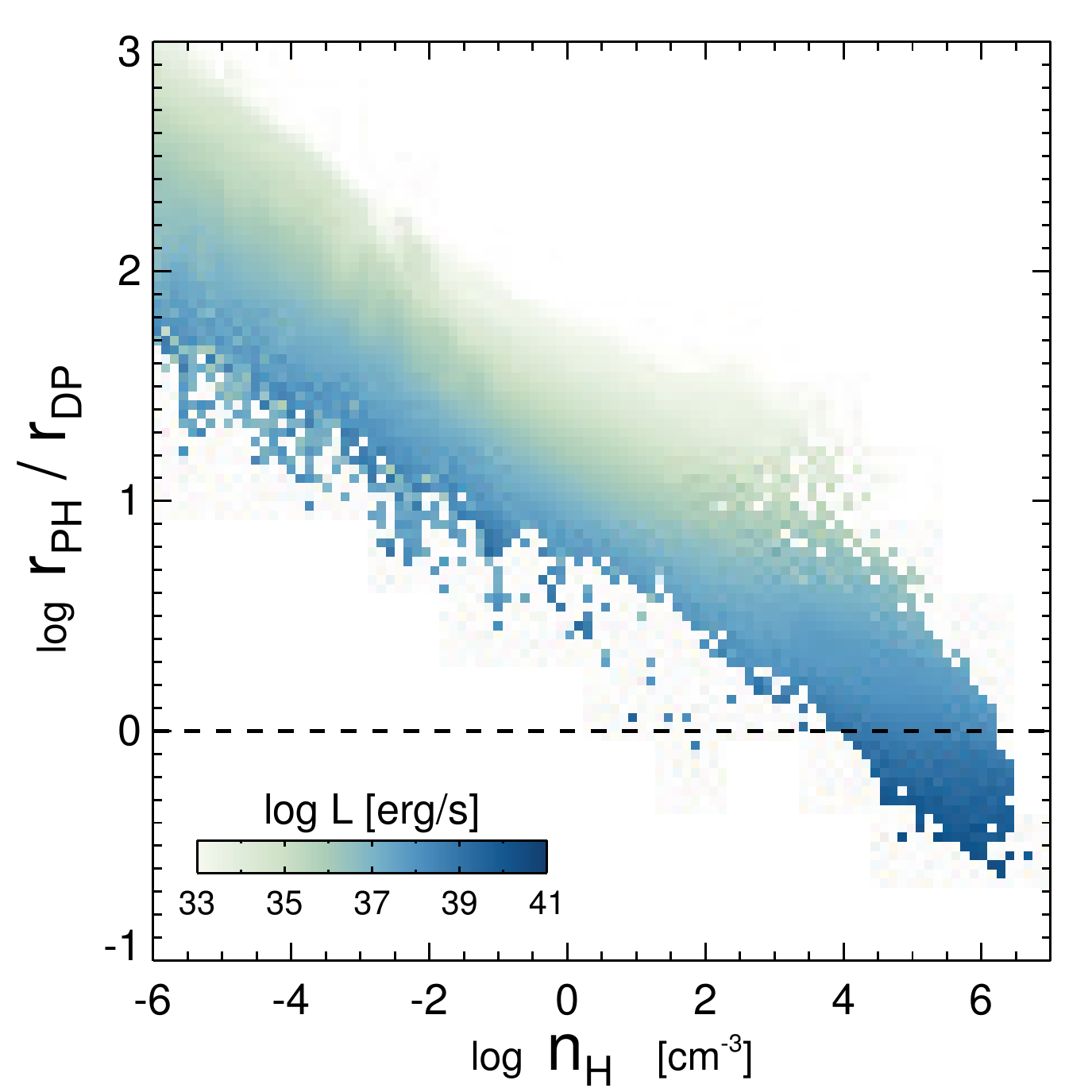} 
   \caption{Comparison between the maximum radius within which photoionisation can 
   counter-balance the external pressure ($r_{\rm PH}$) and the radius within which 
   direct radiation pressure can overcome the external pressure ($r_{\rm DP}$). 
   Only regions cooler than $T\le 2\times10^4\,{\rm K}$ are shown. 
   Darker colours indicate the regions with stronger radiation field.
   The plot suggests that photo-ionisation is more important mechanism than direct 
   radiation pressure in most regions, although direct pressure can be slightly more 
   effective than photo-heating at emptying the gas in very dense regions. }
   \label{fig:mech}
\end{figure}

\begin{figure}
   \centering
   \includegraphics[width=8.0cm]{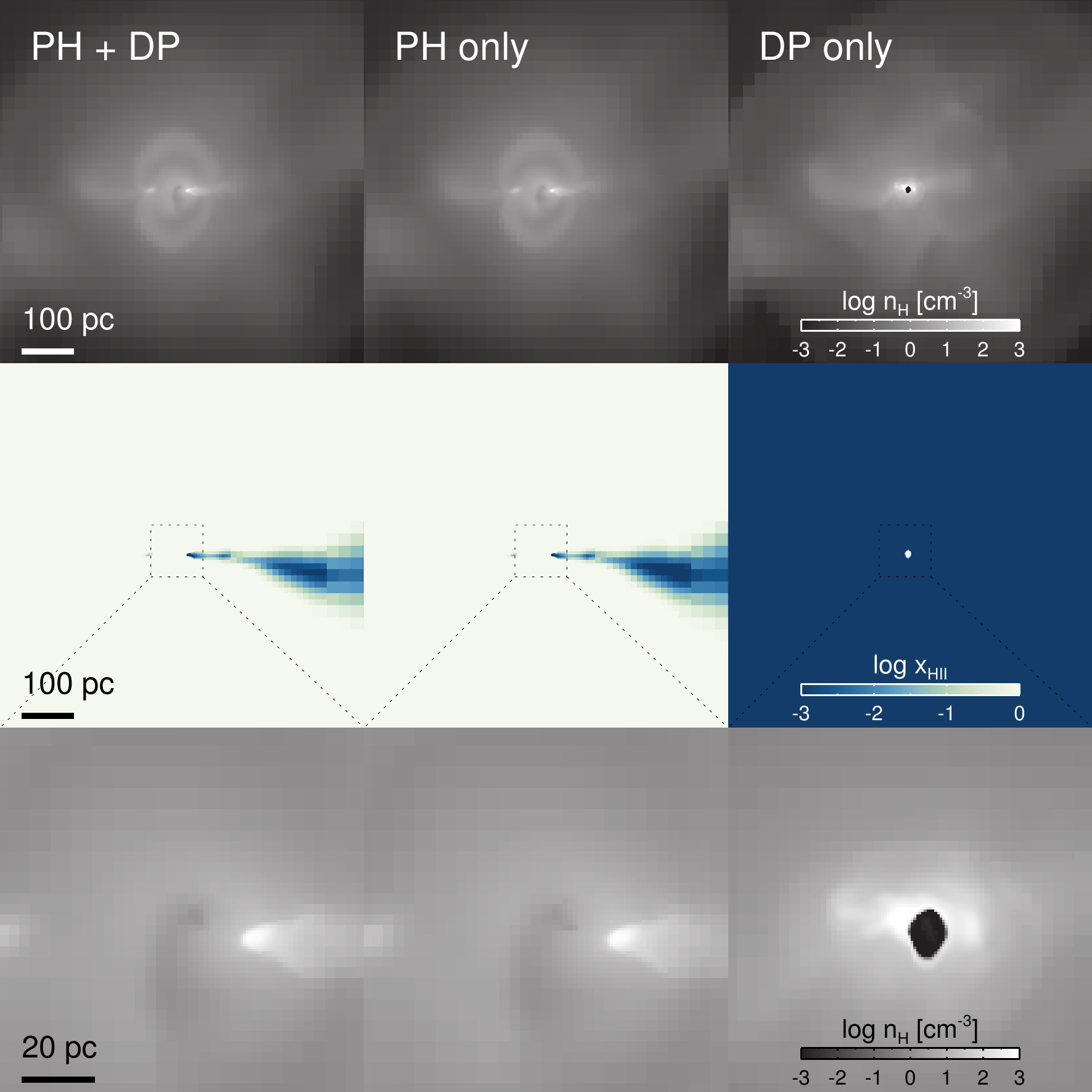} 
   \caption{Effects of photo-heating (PH) and radiation pressure from UV photons (DP). 
   The figure shows a slice of density (first and third rows) and ionised hydrogen fraction 
   distributions (second row) from the simulations with and without PH and/or DP 
   right after a Pop III star with $\approx 150\, \msun$ is formed.
   The virial radius of the mini-halo is $\rvir\sim350\,{\rm pc}$, which is approximately 
   the size of the panels in the first and second rows. The panels in the third row 
   are the zoomed-in images of the panels in the first row. It is evident that photo-heating 
   is responsible for the efficient escape of LyC photons. }
   \label{fig:physics}
\end{figure}

In Figure~\ref{fig:mech}, we compare the two radii given by  Equations~(\ref{eq:rPH}) 
and (\ref{eq:rDP}). We take the external pressure as the mass-weighted pressure in the 
neighbouring 26 cells. This approach is, no doubt, a simplification, given that the immediate 
neighbours may already be ionised and cannot represent the cold surroundings. 
Nevertheless, this is justifiable because we are interested in the mostly neutral 
region where the impact of direct radiation pressure can be strong.
The intensity of ionising radiation is calculated by combining the luminosity of 
each star residing in each cell. Note that we only consider cells with temperatures 
low enough to contain neutral hydrogen ($T \le 2\times10^4\,{\rm K}$).
Figure~\ref{fig:mech} suggests that photo-heating is primarily  responsible for blowing 
gas away in the ISM at densities lower than $\nH\sim10^4\,\cmq$, whereas 
direct radiation pressure dominates over photo-heating at very high densities.
It is worth noting that, at such high densities, pressure equilibrium 
due to direct radiation pressure is achieved on very short time scales 
(i.e. less than a Myr, see Figure~4 in \citealt{rosdahl15a}). Therefore, 
photo-heating is more likely to be responsible for the decrease in gas density 
than direct radiation pressure during the lifetime of massive stars \citep[c.f.,][]{haehnelt95}.

To isolate the effects of photo-ionisation heating and radiation pressure, we identify 
a Pop III forming halo showing a high escape fraction ($\fesc \sim 10\%$) and 
run three simulations varying the included  feedback processes. In the simulation without 
photo-heating (DP only), we compute the non-equilibrium chemistry without increasing 
the temperature by ionisation of hydrogen and helium. Figure~\ref{fig:physics} shows 
slices of density and ionised hydrogen fraction immediately preceding the explosion of a Pop III star 
with $M\approx145\,\msun$. It is evident from this figure that the efficient escape of LyC 
photons can be attributed to heating by photo-ionisation. The central star-forming 
cloud is effectively destroyed only in the runs with photo-heating. 

Interestingly, we find that the run without photo-heating results in $\fesc=0$, 
even though direct radiation pressure creates low-density holes in the 
central region of the dark matter halo. The density of the central hole is even lower than 
the run with photo-heating, because the UV photons accelerate the central gas  
more effectively due to lower temperature in the early stage of the expansion. 
Note that in order to receive a momentum boost from direct radiation pressure,
atoms must be neutral and thus able to absorb ionizing radiation.
Because UV photons from the Pop III star 
propagate and push the gas only in the radial direction, a dense shell structure that 
continually absorbs the ionising radiation is formed. This is an important difference from 
the run which includes photo-heating where the gas becomes over-pressurised and develops 
low-density channels naturally.

\begin{figure}
   \centering
   \includegraphics[width=8.5cm]{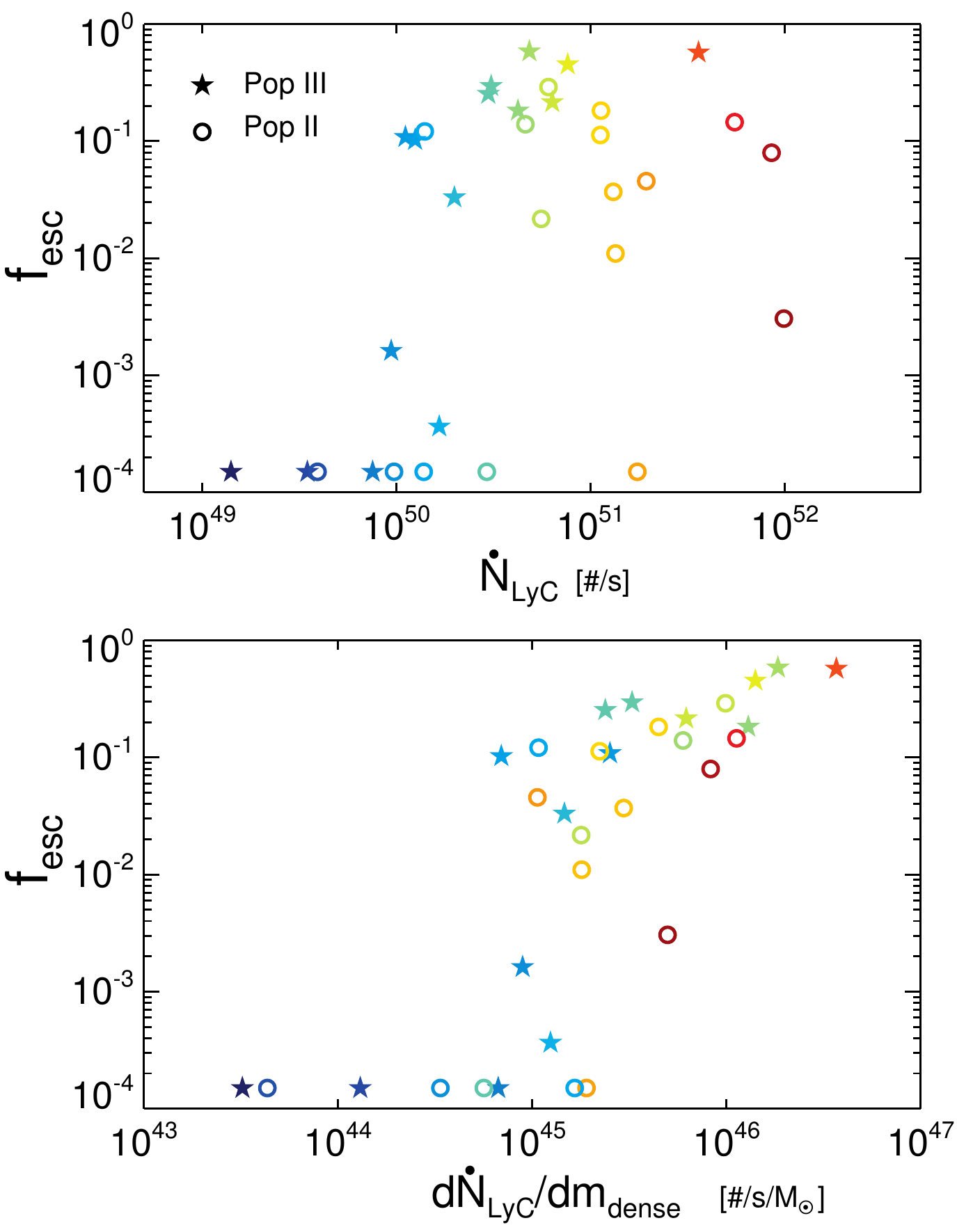} 
   \caption{ Origin of the high escape fraction. 
   {\it Top panel} shows the photon-number weighted escape fraction of each star formation
    event as a function of the peak photon production rate ($\dot{N}_{\rm LyC}$), 
    which is essentially proportional to star formation rate. Each event involving Pop II and 
    III stars is shown with different symbols, as indicated in the legend. Redder colours 
    denote a larger $\dot{N}_{\rm LyC}$. The escape fraction is weakly correlated with 
    $\dot{N}_{\rm LyC}$.
   {\it Bottom panel}: Same as the top panel, but with respect to the peak
   photon production rate per lowest 10 \% dense gas mass from each solid angle
   ($\frac{d\dot{N}_{\rm LyC}}{d\Omega} / \frac{dm_{\rm dense}}{d\Omega}$). Using the lowest 10 \% 
   is motivated by the fact  that LyC photons usually escape through a small opening
   angle with little neutral hydrogen. Here
   we define the dense gas as $n_{\rm H}\ge n_{\rm H,strom}$, where $n_{\rm H,strom}$ 
   is the minimum density below which gas can be fully ionised for a given $N_{\rm LyC}$.
   The escape fraction is higher if there are  more photons available 
   per dense gas mass. We interpret this as being mainly due to photoionisation feedback
   being more effective when there is less dense gas.
   }
   \label{fig:why}
\end{figure}

What, then, determines the escape fraction of LyC photons in the mini-haloes?
Given the importance of photo-heating, the escape fraction is likely to depend on the strength of the 
radiation field.
To investigate this issue, we measure the photon-number weighted escape fraction 
and the peak production rate of LyC photons for individual star formation episodes.
Figure~\ref{fig:why} (top panel) shows that the escape fraction is not strongly 
correlated with the peak production rate of LyC photons, indicating that it is not a simple 
function of star formation rate. However, taking the sample of Pop III stars alone, 
there is a trend that more massive stars show higher escape fractions, 
as seen in Figure~\ref{fig:fesc_pop3}. 
Given that the masses of their host dark matter haloes are more or less the same 
($6.25\la \log \mhalo \la 7$), we hypothesise that the effect of photo-heating 
is likely to be mitigated if stars are enshrouded by a large amount of dense gas.
To test this idea, we measure the gas mass for each solid angle assuming that 
a star cluster is a point source within the virial sphere. We exclude the gas 
less dense than $n_{\rm strom}$, which is the maximum density below which 
recombination can be offset by the photons from Pop III or Pop II clusters.
In addition, since photons in the outskirts of dense clouds are likely to escape more easily and 
LyC photons usually escape through a small opening angle with a low neutral fraction 
\citep[e.g.,][]{cen15,kimj13}, we use the lowest 10\% of the projected mass 
for each solid angle. Note that this is essentially the same as probing when 
the escape fraction is greater than 10\%.  
The bottom panel of Figure~\ref{fig:why} reveals that mini-haloes exhibit a high escape fraction  
when they form a lot of stars while there is a small amount of enshrouding gas.
This is not particularly surprising and is also physically sensible because the more photons 
the halo has per dense gas mass, the more likely the ISM will be ionised and over-pressurised.
One might wonder whether or not the correlation arises simply due to the slim geometry 
of enshrouding gas. We confirm that taking the mean of the dense gas mass for 
each solid angle rather than the lowest 10\% mass does not wash out the trend, 
but only weakens it slightly, suggesting that geometry is a secondary effect.

\subsection{Importance of mini-haloes for reionisation}

Our simulations represent only a small volume of the observable Universe, 
and thus it is not possible to calculate $\tau_e$ directly.
Instead, we take a simple analytic approach to infer the importance of the mini-haloes 
to the reionisation history by computing the evolution of the mass fraction of ionised hydrogen ($Q_{\rm HII}$), 
as
\begin{equation}
\frac{d Q_{\rm HII} }{dt} = \frac{\dot{n}_{\rm ion}}{\left<n_{\rm H}\right>} - \frac{Q_{\rm HII}}{t_{\rm rec}(C_{\rm HII})},
\label{eq:reion}
\end{equation}
where $\left< n_{\rm H}\right>$ is the mean density of the universe in comoving scale.
The recombination timescale ($t_{\rm rec}$) is a function of a clumping factor ($C_{\rm HII}$) and temperature, as 
\begin{equation}
t_{\rm rec} (C_{\rm HII})= \left[ C_{\rm HII}\,\alpha_{\rm B}(T)\, f_e(z)\, \left<n_{\rm H}\right> \,(1+z)^3\right]^{-1}
\end{equation}
where a correction factor ($f_e(z)$) is included to account for the additional contribution from 
singly ($z>4$) or doubly ($z<4$) ionised helium to the number density of electron \citep[e.g.,][]{kuhlen12}. 
We adopt a redshift-dependent "effective" clumping factor of $C_{\rm HII} = 1 + \exp(-0.28\, z +3.59)$ at $z\ge10$ or
$C_{\rm HII} = 3.2$ at $z<10$ \citep{pawlik09}.
Once $Q_{\rm HII}$ is determined, the Thomson optical depth is calculated as 
\begin{equation}
\tau_e (z)= c \left<n_{\rm H}\right>\,\sigma_T\, \int_0^z f_e(z)\,Q_{\rm HII}(z') \frac{(1+z')^2 dz'}{H(z')}.
\end{equation}

\begin{figure*}
   \centering
   \includegraphics[width=8cm]{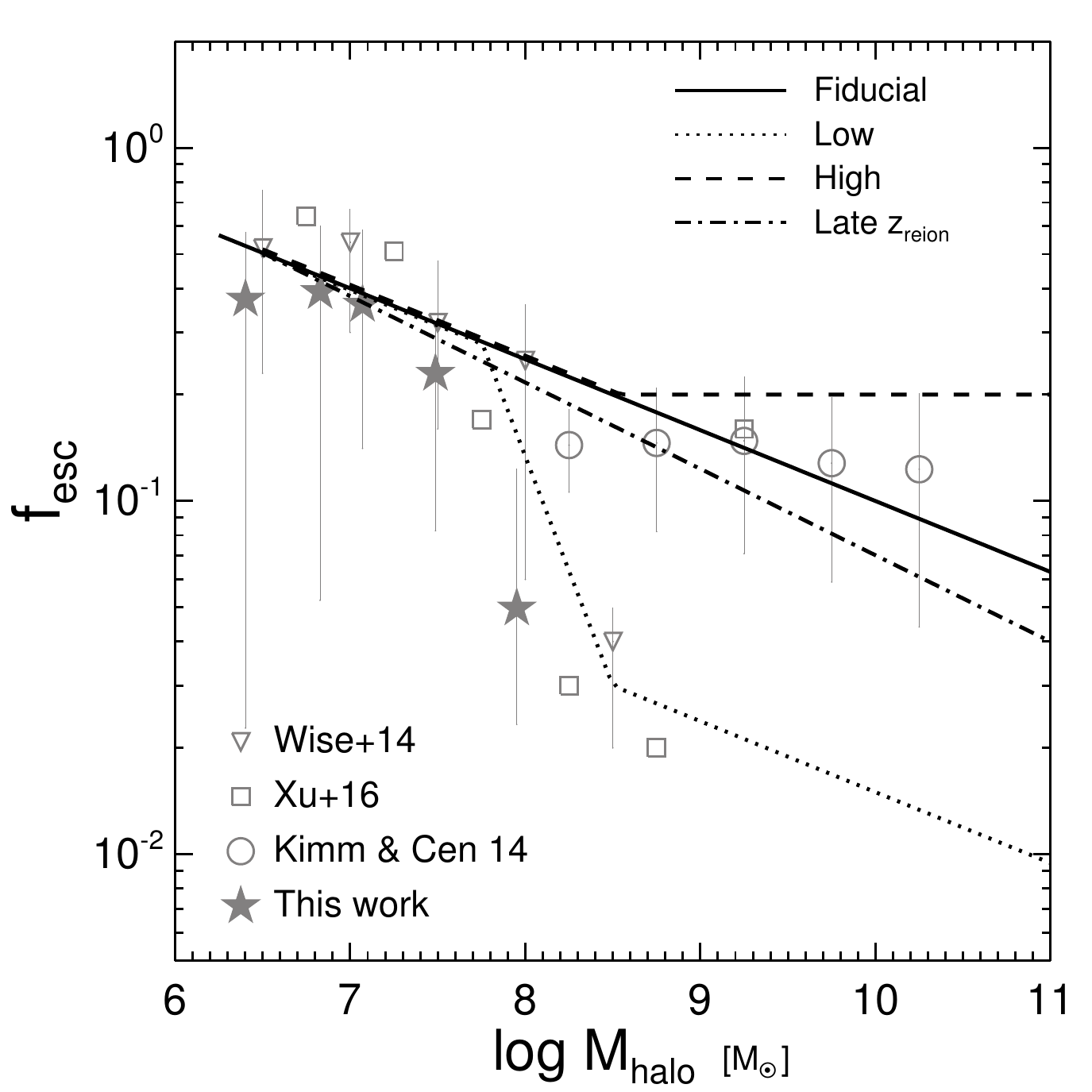} 
  \includegraphics[width=8cm]{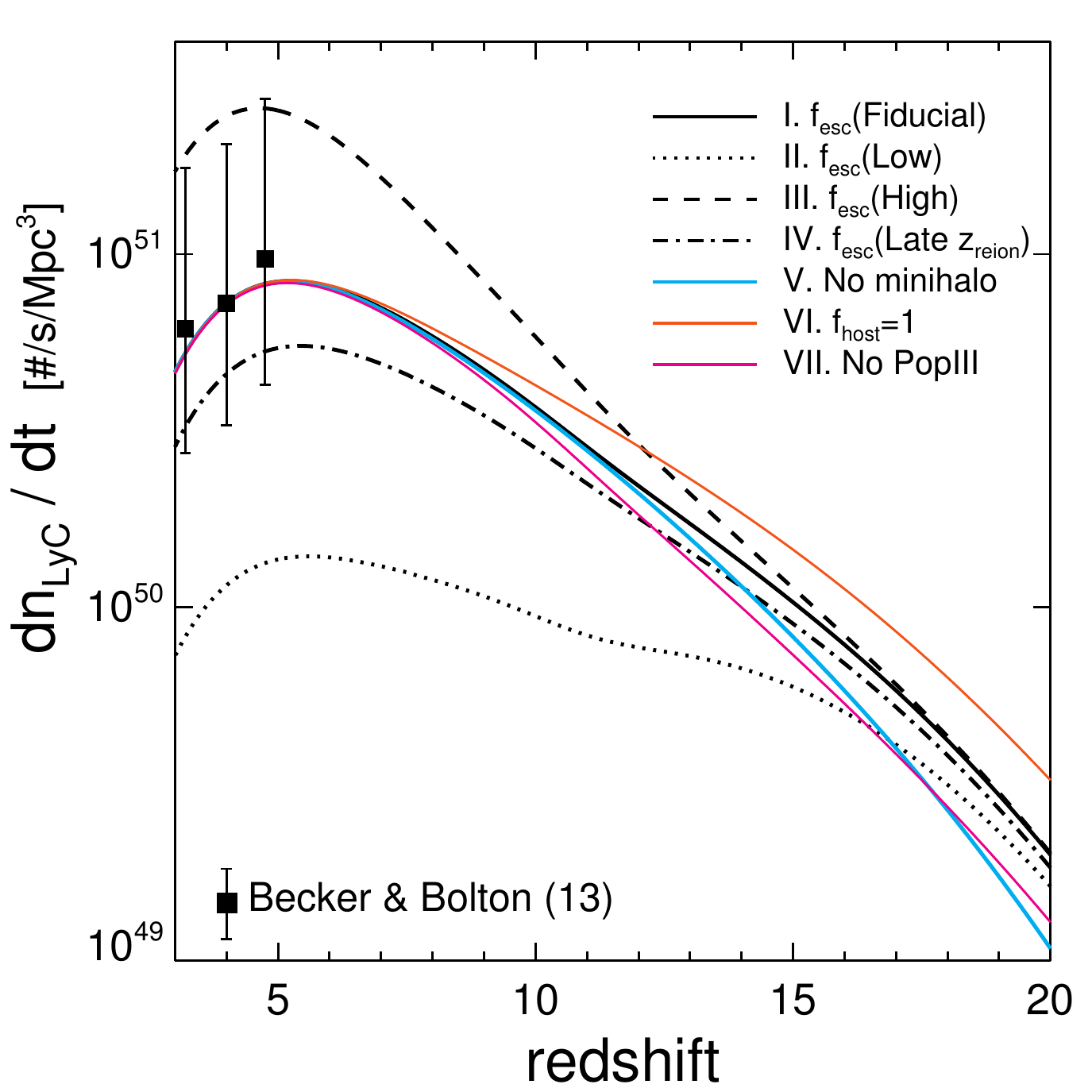} 
   \includegraphics[width=8cm]{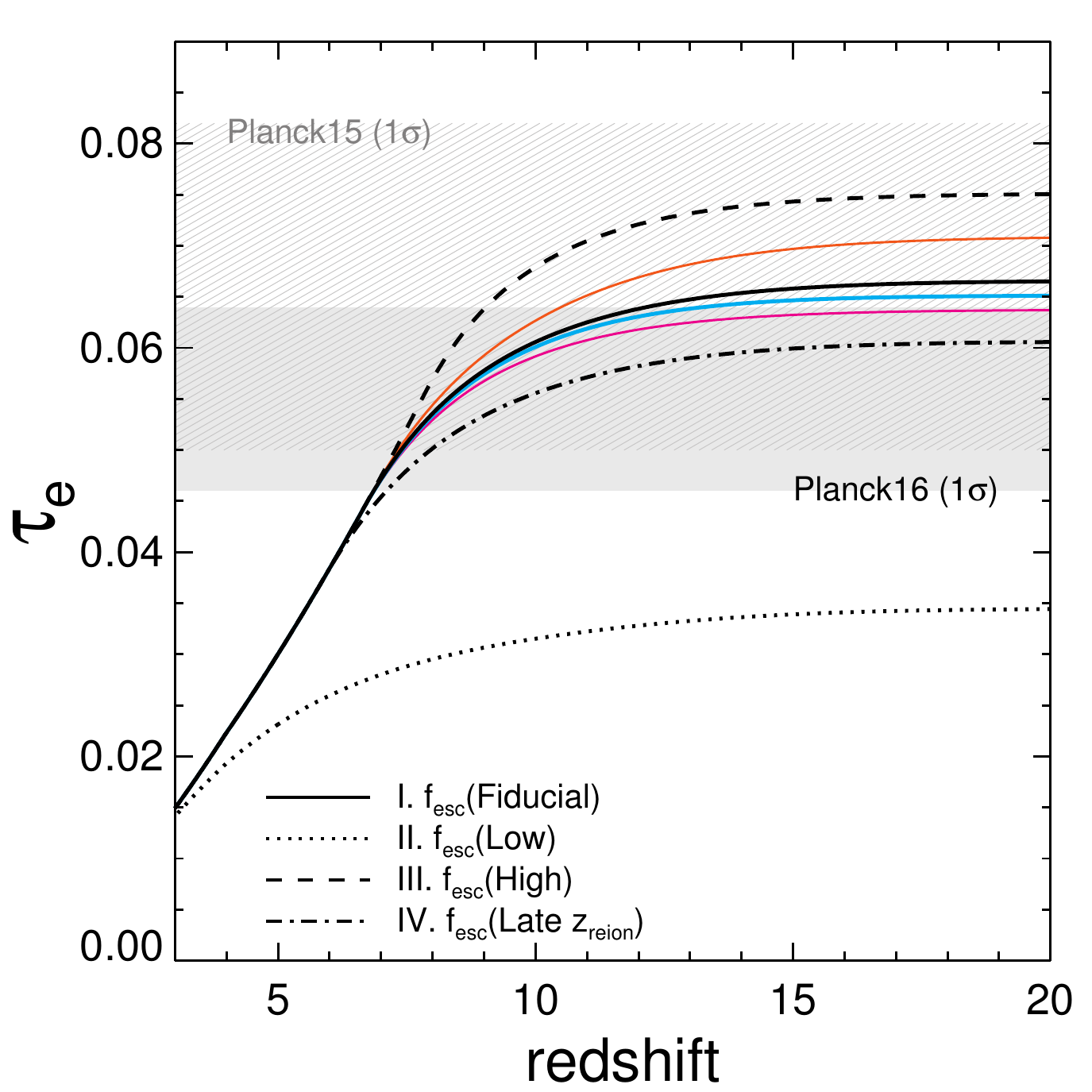} 
   \includegraphics[width=8cm]{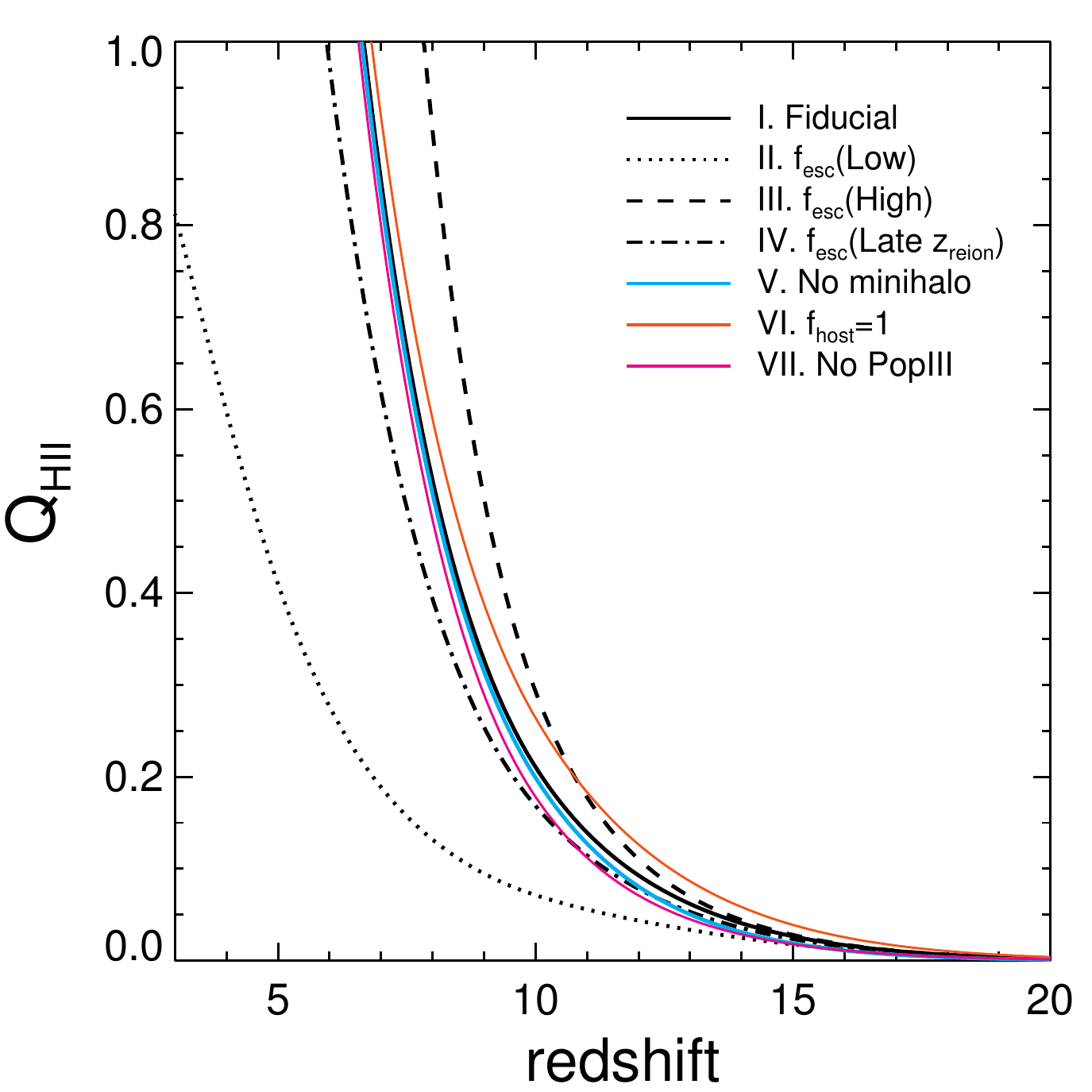} 
   \caption{
   Reionisation models based on simple analytic calculations.
   We examine three different escape fractions (models I, II, and III), the presence of the
   mini-haloes (model V) or Pop III stars (model VII), and the occupation fraction 
   of the mini-haloes (model VI). Also included as model IV is the escape fraction favoured
   by the patchy reionisation picture \citep{becker15}.
   {\it Top left}: different assumptions about the escape fraction used to 
   calculate the reionisation history. Different symbols indicate the escape fraction 
   from different theoretical studies, as indicated in the legend. The solid, dashed, and dotted 
   line represent the fiducial, high, and low escape fraction model, respectively. 
   {\it Top right}: The photon production rate per unit volume ($dn_{\rm LyC}/dt$) as a 
   function of redshift. 
   We use the \citet{sheth02} mass functions and our theoretical results (the dashed line in 
   Figure~\ref{fig:mstar}, Equation~\ref{eq:mstar}) to compute $dn_{\rm LyC}/dt$. 
   {\it Bottom left:} The electron optical depth ($\tau_e$) in different models. 
   The optical depths inferred from the Planck experiments are shown as shaded regions
   (light grey: \citealt{planck-collaboration15}, dark grey: \citealt{planck-collaboration16}).
   {\it Bottom right:}  The fraction of ionised hydrogen.
   Our fiducial model predicts the reionisation redshift of 6.7 and $\tau_e=0.067$.
   We find that the mini-haloes produce a rather small  amount of LyC photons,
   and thus are of minor importance for  the reionisation of the Universe.}
   \label{fig:tau}
\end{figure*}

The most crucial term in Equation~(\ref{eq:reion}) is the photon production rate 
per unit volume $\dot{n}_{\rm ion}$ (${\rm \#\, Mpc^{-3}\,s^{-1}}$). 
To compute this quantity, we first generate dark matter halo mass functions 
as a function of redshift \citep{sheth02}, and assign stars to each dark matter halo 
by taking into account the following factors:
\begin{itemize}
\item[a)] {\it Star formation efficiency} \\ 
\citet{kimm14} performed cosmological radiation-hydrodynamic simulations, and 
showed that the UV luminosity function at $z\sim7$ can be reproduced with efficient 
SN feedback. They find that the stellar mass to the dark matter halo mass
may be approximated by the fit, 
\begin{equation}
\log \mstar = -8.08 + 1.55\,\log \mhalo
\label{eq:mstar}
\end{equation}
(see Figure~\ref{fig:mstar}, the black line). Our simulations suggest that the relation 
can be extended even to the mini-halo regime at $7\la z \la 15$. 
Thus, we use this fit to compute the total stellar mass formed in each dark 
matter halo assuming no redshift dependence.
 \\

\item[b)] {\it Occupation fraction} \\
Not all of the mini-haloes hosts Pop II stars. This is essentially because early stars 
pre-heat the IGM and suppress the accretion on to the 
halo \citep{gnedin00}. Moreover, the formation of molecular hydrogen, which is 
a prerequisite for gas collapse in mini-haloes in the absence of metals, 
is suppressed by the Lyman-Werner radiation from the neighbouring haloes. 
Furthermore, gas in haloes can be blown out by efficient feedback processes 
from Pop III stars. To account for the occupation 
fraction in haloes with $\mhalo\ge10^7\,\msun$, 
we use a redshift-dependent fit to the fraction of haloes that host stars in the \citet{wise14} simulations (see their Figure~2). This corresponds to 
$f_{\rm host}\approx1$ for $\mhalo \ge 10^7\,\msun$ at $z\ga17$, while $f_{\rm host}$ drops to less than 0.01 in haloes with $10^7\,\msun$ at $z\sim7$. 
For haloes with $\mhalo\le10^{6.25}\,\msun$ and $\mhalo\ge10^8\,\msun$, we adopt a fixed 
value of $f_{\rm host}=2.4\times10^{-4}$ and  unity, respectively \citep{wise14}. 
The occupation fraction is considered 
by simply reducing the number density of dark matter haloes that can form stars
in estimating $\dot{n}_{\rm ion}$  (Equation~\ref{eq:reion}). \\

\item[c)] {\it Escape fraction} \\
Our estimates of the escape fraction in mini-haloes are in good agreement with the results 
performed with {\sc enzo}  \citep{wise14,xu16}. However, the escape fractions in 
atomic-cooling haloes appears to be less certain, and thus we compare 
three different models:
i) an escape fraction that slowly decreases with the halo mass \citep{kimm14,xu16}, 
ii) an escape fraction showing a more rapid decline \citep[e.g.,][]{paardekooper15,ma15}, and
iii) an escape fraction with a lower limit of 20 \%.
For the fiducial case, we employ the fit,  
\begin{equation}
\log \fesc = 1.0 - 0.2 \,\log \mhalo,
\end{equation}
whereas a broken law is adopted for the ``low'' case
\begin{equation}
\log \fesc =\left\{
\begin{array}{ll}
 0.17 - 0.2\,\log \mhalo &  (\mhalo > 10^{8.5}) \\
 9.52 - 1.3 \,\log \mhalo &  (10^{7.75} < \mhalo \le 10^{8.5}) \\
1.0 - 0.2 \,\log \mhalo &  ( \mhalo \le 10^{7.75})\\
\end{array} 
\right. .
\end{equation}
The last model has the same escape fraction as the fiducial case, 
but with the minimum of $\fesc=20\%$, as is often assumed in reionisation studies.
The three models are shown in Figure~\ref{fig:tau} (the top left panel). 
Also included as a dot-dashed line is a model with an  escape fraction that would be necessary 
to yield a later end of reionisation as perhaps suggested by the \lya\ opacity data presented by \citet{becker15}. 
\begin{equation}
\log \fesc = 1.292 - 0.245 \,\log \mhalo,
\end{equation}
\\

\item[d)] {\it Metallicity-dependent $\dot{N}_{\rm ion}$}\\
Metal-rich main sequence stars are cooler than metal-poor stars,
and because of the larger opacity in the stellar atmosphere
a SSP of solar metallicity produces approximately a 
factor of 2.5 fewer LyC photons than a SSP with $0.02\, Z_\odot$.
In order to not over-estimate the photon budget from metal-rich, massive 
galaxies, we take into consideration the metallicity-dependent $\dot{N}_{\rm ion}$
by using the local mass-metallicity relation for dwarf 
galaxies \citep{woo08}, as $\log Z = -3.7 + 0.4\, \log \left(\mstar/10^6\,\msun\right)$.
Although this is certainly a simplification that needs to be tested against upcoming 
high-z observations, we note that our simulated galaxies are metal-poor  
by about a factor of two compared to the relation, well within the scatter of the measurements.
Once the metallicity of stars is determined, we use the metallicity-dependent number of 
LyC photons per $\msun$, as
\begin{equation}
\log Q_{\rm LyC} = 60.31 - 0.237 \, \log Z ,
\end{equation}
which is appropriate for the Padova AGB models with a Kroupa IMF \citep{leitherer99}.
Here we adopt  a maximum mass cut-off of 120 $\msun$ for the metallicity range $0.0004 \le Z \le 0.05$. 
We do not extrapolate $Q_{\rm LyC}$ for metallicities lower than $0.02\,Z_{\odot}$. 
It is worth noting that for  $Z=0.004$,  
the Padova AGB model predicts $Q_{\rm LyC}=8.86 \times10^{60}$,
while the binary evolutionary model with  a 300 $\msun$ cut off gives a slightly higher 
estimate of $Q_{\rm LyC}=9.26\times10^{60}$ \citep{stanway16}.
Stellar models with strong rotation seem to generate even more photons 
$Q_{\rm LyC}=11.2\times10^{60}$ at $Z=0.002$ \citep{topping15},
but this leads to only a minor increase ($\sim$ 20 \%) in the LyC production rate, 
which does not change our conclusions significantly.
\\

\item[e)] {\it Pop III stars} \\
The largest uncertainty is perhaps 
the contribution of Pop III stars to reionisation. Not only is the IMF of the Pop III stars 
not well constrained, but also only a handful of Pop III stars are simulated in this work 
to estimate  their escape fraction.
Thus, we examine two extreme cases, one in which Pop III stars do not contribute 
to reionisation at all, and the other in which 40\% of the photons from Pop III stars with 
$M_{\rm PopIII}\ga100\,\msun$ are assumed to contribute to reionisation. 
The former can be regarded as the case where only low-mass Pop III stars with 
$M_{\rm PopIII}\la100\,\msun$ form, whereas the latter corresponds to the 
case where one allows for the wide range of masses from 10 to 1000 $\msun$. 
Note that the assumption of the mass-dependent escape fraction is motivated by 
Figure~\ref{fig:fesc_pop3}. The total number of LyC photons that Pop III stars can 
generate is computed by convolving the IMF (Equation~\ref{eq:sf_pop3}) and 
the escape fraction. This yields 
$\left< N_{\rm LyC, PopIII}\right> = 1.8\times10^{62} \,\msun^{-1}$. 
We then adopt the formation rate of Pop III from the large volume simulations of \citet{xu16} 
(the ``Normal'' run) by using the fit, 
$\log \dot{M}_{\rm popIII} / (M_\odot\, {\rm yr^{-1}}\,{\rm Mpc^{-3}})= -6.428 + 0.275\,z - 0.011 z^2$. 
This rate is roughly an order of magnitude smaller than the rate employed in
 \citet{wise14}, which is based on the simulation of an over-dense region (``Rare Peak'', H. Xu, priv. comm.).
\end{itemize}

\begin{figure*}
   \centering
   \includegraphics[width=8cm]{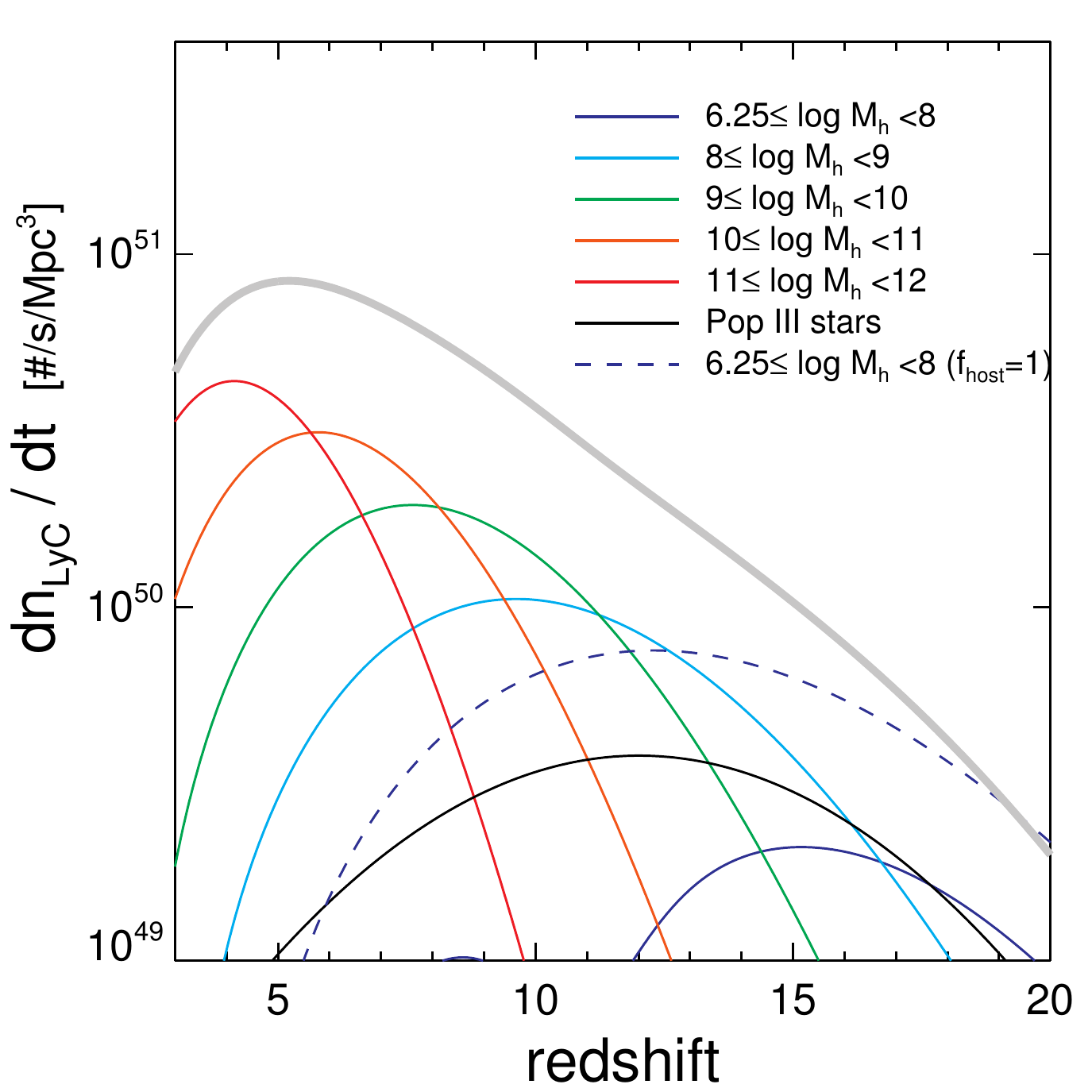} 
  \includegraphics[width=8cm]{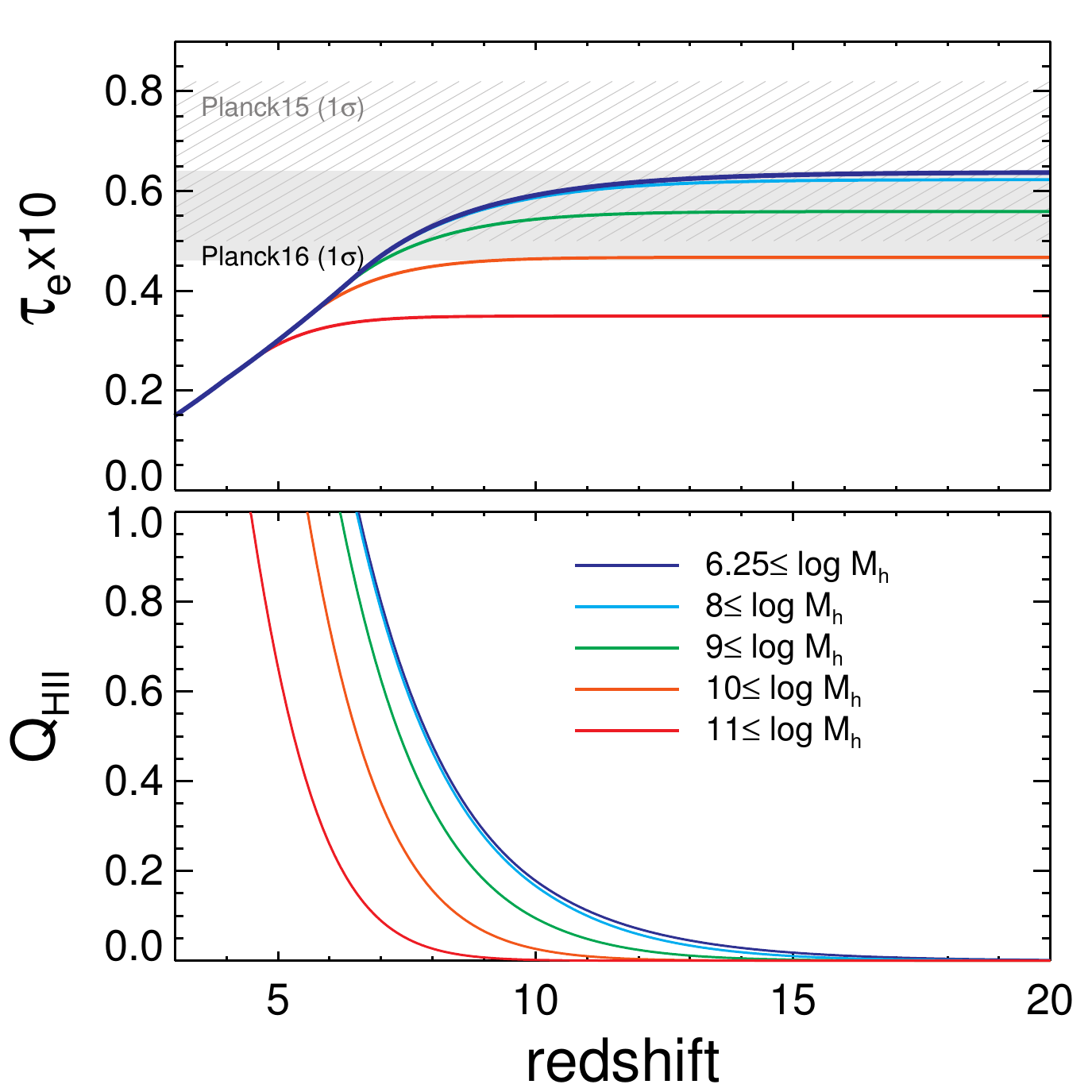} 
   \caption{
   {\it Left}: Contribution of photons from different halo masses to reionisation as a function 
   of redshift. We adopt the fiducial model for escape fractions 
   (the black solid line in Figure~\ref{fig:tau}) and the stellar mass-to-halo mass 
   relation from this work (the dashed line in Figure~\ref{fig:mstar}).
   The photon production rate density from Pop III stars is shown as the black dashed line.
   The dotted line displays the maximum contribution from mini-haloes assuming that 
   all of them host stars ($f_{\rm host}=1$).
   {\it Right:} The electron optical depth (top) and the fraction of ionised 
   hydrogen (bottom) for models with different minimum halo masses, as indicated 
   in the legend. The plots demonstrate that it is necessary to resolve haloes of mass 
  $\approx10^8\,\msun$ to model reionisation in simulations. }
  \label{fig:tau_break}
\end{figure*}

Once the number of photons escaping from a dark matter halo ($N_{\rm LyC}^{\rm esc}$) 
as a function of $\mhalo$ and $z$ is computed, the total number of escaped photons 
until $z$ can be obtained by multiplying the dark matter halo mass function 
with $N_{\rm LyC}^{\rm esc}$.  
We use the \citet{sheth02} mass function for the mass range  
$\mhalo=10^{6.25}\,\msun$ to $10^{12}\,\msun$. The time derivative of the integrated quantity 
corresponds to $\dot n_{\rm ion}$ in Equation~(\ref{eq:reion}).

The top right panel in Figure~\ref{fig:tau} shows the evolution of $\dot n_{\rm ion}$ 
for the different models. In the cases with the fiducial escape fraction, the number of LyC 
photons that escape from their host halo peaks at $z\sim5$ and decreases at lower redshift,
in remarkably good agreement  with the observational findings at $3\le z \le 4.7$ \citep{becker13}.
Note that the peak is earlier than the peak of the star formation rate density 
\citep[$z\sim3$, e.g.][]{hopkins06}, because the photon budget at $z\la6$ is 
determined by more massive, metal-enriched galaxies where  fewer
ionising photons are  generated per mass of stars  formed than in  dwarf galaxies.
The peak would  shift to even earlier redshifts 
if  the escape fraction  decreases at lower redshifts. 
Although we assume a redshift-independent escape fraction, which is motivated 
by our simulations and simulations of Wise and collaborators at $z\ga7$,
there is a hint that the escape fraction decreases with decreasing redshift.
For example, from cosmological radiation-hydrodynamic 
simulations, \citet{cen15} measured that the escape fraction is decreased by a factor of $\sim 2$ at $z\sim4$, 
compared to galaxies at $z\ga7$ \citep{kimm14}. They attribute this to the fact 
that birth clouds are more slowly disrupted, as star formation becomes less stochastic 
at lower redshift. Thus, we caution the readers that $\dot{n}_{\rm ion}$  at $z \le 6$ is probably still 
uncertain by a factor of a few. Our  redshift-independent assumption should thus be kept in mind 
in our discussion of the  reionisation  history of the Universe  at $z\ga6$.

Figure~\ref{fig:tau} depicts the evolution of the electron optical depth 
($\tau_e$), and the fraction of ionised gas ($Q_{\rm HII}$). 
Several important conclusions can be drawn from this plot.  First, the mini-haloes 
turn out to be of minor importance to reionisation. 
Although their escape fraction is high, the photon production rate is intrinsically low,
firstly because stellar feedback efficiently suppress star formation,
and secondly because not every mini-halo hosts stars \citep{gnedin00,wise14}.
We find that even if all the mini-haloes are assumed to host stars, 
$\tau_e$ will only increase slightly (model VI). Conversely, if we neglect the mini-haloes, 
there is no noticable difference to the prediction of the reionisation history.
This is in stark contrast with the claim that mini-haloes provide enough photons 
in the early universe to achieve a high $\tau_e\ga0.1$ \citep{wise14}. 
The difference can  be attributed to a large part  to the reduced stellar masses in the 
smallest mini-haloes in this work, compared to  \citet{wise14}.

Similarly, we find that the contribution from Pop III stars is not substantial 
\citep[see also][c.f., \citealt{ahn12,maio16}]{paardekooper13}. 
The model without Pop III stars (model VI) predicts a minor decrease 
(35\% and 10\%) in the number of ionising photons at $z\sim20$ and $z\sim10$, 
respectively (see also Figure~\ref{fig:tau_break}). The resulting difference in $\tau_e$ is 
found to be much smaller than the 1 $\sigma$ error of the Planck measurements. 
This strongly suggests that even if the characteristic mass of Pop III stars is 
lower ($\sim40\,\msun$), which prohibits them from ionising 
the IGM, the optical depth measurement may not be able to 
put strong constraints on the properties of Pop III stars \citep[c.f.,][]{visbal15}.

Instead, we find that the prediction of $\tau_e$ is primarily sensitive to the escape fraction 
of the atomic cooling haloes ($\mhalo\ga10^8\,\msun$). 
Our fiducial model with the relatively high escape 
fraction and Pop III stars yields $\tau_e \approx 0.067$, which is in reasonable 
agreement with the Planck measurements \citep{planck-collaboration15,planck-collaboration16}.
We also find that the evolution of $Q_{\rm HII}$ lies between that of the `Late' and the `Very Late' 
reionisation model of \citet{choudhury15} at $z\ge8$ \citep[see also][]{kulkarni16}, 
both of which are shown to be able to reproduce the rapid drop 
in the fraction of \lya\ emitters at $z\ga7$ \citep{choudhury15}.
On the contrary, the model with 
the "Low" escape fraction in the atomic-cooling regime  fails to 
ionise the universe by $z\sim3$ and under-predicts $\tau_e$, even in the presence of 
the mini-haloes and Pop III stars. This demonstrates that massive stars in  
atomic-cooling haloes are likely to be the main sources responsible for reionising the Universe. 
In order for this ``Low'' escape fraction model to fully ionise the Universe by $z\sim6$, 
a large number of AGN would be necessary at high redshifts to provide 
enough LyC photons \citep[e.g.,][]{madau15}. This scenario is certainly not yet ruled out,
but it depends critically on the assumption about the evolution of the emissivity
as a function of redshift, which is still a matter of debate \citep{giallongo15,kashikawa15,kimy15,chardin15}.

When the escape fraction is assumed to always be greater 
than 20\%, the optical depth is predicted by our models to be rather high ($\tau_e=0.075$) 
and reionisation ends early ($z\sim8$), which is less favoured by 
the latest Planck measurements ($\tau_e=0.055\pm0.009$). Such a model also struggles 
to explain the evolution of the \lya\ opacity (fluctuations) in QSO spectra at $z>5$ 
as well as the rapid evolution of \lya\ emitters \citep{becker15,chardin15,choudhury15}. 
In fact, even our fiducial model predicts reionisation 
at $z\sim6.7$, slightly earlier than in the fiducial model of \citet{chardin15} that fits well the 
photo-ionisation rate inferred from \lya\ forest data 
and perhaps suggesting that the escape of LyC photons in the atomic-cooling haloes needs 
to be suppressed further.

It should be noted, however, that there is a possibility that the simple analytic calculations 
are likely to predict too rapid propagation of HII bubbles in the late stage of reionisation ($z\la8$)
due to the assumption of infinite speed of light used in Equation~(\ref{eq:reion}).
In practice, photons from massive haloes will have to travel to encounter neutral hydrogen in 
regions devoid of sources, which inevitably delays the propagation of HII bubbles.
If this is the case, reionisation in the fiducial model is likely to end later than $z=6.7$, 
bringing the optical depth in better agreement with the latest Planck results.  
However, the precise determination of the delay requires large-scale reionisation simulations 
that are calibrated to match the photon production rates from haloes of different masses in this study, 
hence it is not clear yet how significant the effect will be.
If the propagation of HII bubbles is not significantly overestimated in
our simple calculations, 
the escape fraction needs to be reduced further in more massive haloes in order for dwarf galaxies 
to reionise the universe only by $z\le6$, 
as in our ``Late $z_{\rm reion}$'' model shown by  the dot-dashed line in Figure~\ref{fig:tau}. 
Such escape fractions in atomic-cooling haloes are  certainly  within the uncertainty of current numerical results, 
and it would be worthwhile to re-examine escape fractions in more massive 
haloes ($\mhalo \ge 10^{10}\,\msun$) with the physically well-motivated 
thermo-turbulent star formation model employed here where stars form preferentially in 
gravitationally bound regions, and see how it compares with the results 
based on a simple density criterion for star formation \citep{kimm14}.

In the left panel of Figure~\ref{fig:tau_break}, we show the relative contribution of LyC photons 
from different components for our fiducial reionisation model. It can be seen that more massive 
haloes dominate the 
photon budget at lower redshift in this model, and that reionisation is driven by intermediate-mass 
atomic cooling haloes ($10^8\,\msun \la \mhalo\la10^{11}\,\msun$).
The current theoretical estimate of  
Pop III star formation rates \citep{xu16} suggests that Pop III stars are 
important only at $z\ga15$. The plot also demonstrates that the photon 
production rate density decreases at $z \le 5$ in our model, 
essentially because the number density of the intermediate-mass haloes with 
$10^8\,\msun \la \mhalo\la10^{11}\,\msun$ does not increase 
anymore  at lower redshift. One may wonder here whether the inclusion of more massive
haloes with masses  $\mhalo\ga10^{12}\,\msun$ would change our conclusions, but we find that 
the photon production rate density from these haloes is significantly smaller than 
that from $ \mhalo \ge 10^{11}\,\msun$ at $z\ge3$ 
if the local $\mstar$--$\mhalo$ relation \citep{behroozi13} is assumed,
as star formation becomes less efficient in haloes with $\mhalo\ge10^{12}\,\msun$.

Our results indicate that it will be necessary to resolve  haloes of mass 
$\approx10^8\,\msun$ in order to correctly capture the expansion 
of ionised HII bubbles in large-scale simulations.
The right panels of Figure~\ref{fig:tau_break} show that neglecting the contribution of LyC photons
from haloes less massive than $10^9\,\msun$ will only marginally delay the expansion, 
reasonably reproducing the electron optical depth. 
However, simulations that cannot resolve haloes with 
$\mhalo\la 10^{10}\,\msun$ would need to 
adopt higher escape fractions than the values derived in this work 
to provide photons early enough \citep[e.g.,][]{ocvirk15,gnedin14,pawlik15}.

\subsection{Caveats}
\label{sec:caveats}
Even though we include the most important physical ingredients in our simulations, 
it should be mentioned that several potential issues which can affect the predictions 
of star formation and the escape fraction are neglected. 

First, we do not take into account the fact that the mass of the most massive star 
within a cluster is correlated with the total cluster mass \citep{weidner10,kirk11}. 
Given that the most massive stars are the most efficient at producing LyC photons,
the number of photons per unit stellar mass (i.e. specific number) is 
likely to be over-estimated in the simulated small Pop II clusters of mass 
$M\le10^4\,\msun$ where the IMF would not be fully sampled 
\citep[see][for a review]{kroupa13}. 
For example, at $Z=0.0004$, the specific number 
of LyC photons from a $10^3\,\msun$ cluster is smaller by $\sim 60\%$ than that from a 
$10^4\,\msun$ cluster. Neglecting this IMF sampling issue may have suppressed 
star formation more effectively than it should have. However, we do not expect that 
the escape fraction would be affected significantly by lowering $N_{\rm LyC}$ by a 
factor of two, given that the number of Ly$\alpha$ photons per dense gas mass 
spans three orders of magnitude already (Figure~\ref{fig:why}).
Furthermore, it should be noted that the grids of the stellar tracks do not extend to values 
lower than 0.02 $Z_\odot$. For comparison, star particles with $M<10^4\,\msun$ 
in our simulations often have the metallicity of $10^{-4}$--$10^{-3}\,Z_{\odot}$. 
We use the lowest grid point 
in order to not extrapolate the number of photons produced, but the use of 
more appropriate stellar grids is likely to enhance the Ly$\alpha$ photon 
production, possibly compensating the over-estimation due to the incomplete IMF sampling.  

Second, we neglect the contribution from runaway stars, which are thought to form 
through three-body interactions within clusters \citep[e.g.,][]{gies86,leonard88,fujii11} 
and/or the ejection in a binary system after the explosion of the primary 
star \citep[e.g.,][]{blaauw61,portegies-zwart00}. Using an analytic model of disc galaxies, 
\citet{conroy12} claim that the inclusion 
of the runaway stars can in principle enhance the escape fraction by up to an order of 
magnitude if $\fesc$ without them is very small. However, \citet{kimm14} compare the 
radiation-hydrodynamic simulations with and without the runaways, 
and show that the increase in $\fesc$ due to runaways is small ($\sim 20\%$ level) 
mainly because $\fesc$ is already high ($\fesc\sim10\%$). In the same context, 
we do not expect that the runaways significantly affect the prediction of $\fesc$ 
in mini-haloes where $\fesc$ is even higher ($\sim20-40\%$).

Third, our predictions are subject to the stellar evolutionary models.
We take the production rate of LyC photons from the Padova model \citep{girardi00,leitherer99},  
which is based on the evolution of single stars  without rotation. However, it is known that  
many, if not all, massive stars live  in  binaries \citep[e.g.,][]{sana12}, and 
that their spin is non-negligible.
The interactions of binary stars can not only increase the total number of LyC photons, 
but also make the decline of the production rate slower,
as the primary star transfers gas to the secondary, removes the hydrogen envelope,
and the stars merge \citep{stanway16}. Rotation can also increase the total number of ionising photons 
and the lifetime of massive stars by regulating the mixing and fuelling of gas 
into the stellar core \citep[e.g.,][]{ekstrom12}. As aforementioned, the adoption of the 
recent stellar models can increase the production rate of LyC photons 
by $\sim10-20\%$ for an SSP \citep{topping15,stanway16}, and  is therefore 
not very important for the 
total photon production. However, these effects can potentially 
increase the escape fraction to the level of $\sim20\%$ in galaxies hosted 
by massive haloes with $10^{10}$--$10^{11}\,\msun$ \citep{ma16},
as stars older than $t\ga {\rm 3\, Myr}$ produce a non-negligible fraction 
of LyC photons. This could potentially  bring the results of the simulations  in tension with the 
rather late end of reionisation suggested by the CMB data as wells as \lya\ forest 
and \lya\ emitter data, as dwarf galaxies would produce too many photons (see Figure~\ref{fig:tau}).  
Future radiation-hydrodynamic simulations with the self-consistent modelling of binary populations are 
necessary to better understand the role of (massive) binary stars and to see if theory 
and observations can still be reconciled.

Finally, due to the limited computational resources we  had available, 
the work presented here lacks the systematic investigation of the effects of numerical resolution on the prediction of the escape 
of LyC photons. However, we note that our resolution and refinement strategy 
allows for star formation in sufficiently dense and gravitationally bound regions,
which should  be essential to not over-estimate the escape fraction \citep[e.g.,][]{ma15}. 
Nevertheless, in order to make sure that this is really the case, we perform a 
resolution test by re-running the H1 halo with one fewer level of refinement 
(i.e. 1.4 pc resolution). We find that the escape fraction in the mini-halo is still high 
($\fesc\sim40\%$), although the stellar mass is increased by 30\%. 
We also note that the  recent work by \citet{xu16} shows that their 19 pc (comoving) resolution 
simulations ($\sim$ 2 pc, physical) yield the results consistent with their previous simulations 
with  a higher resolution of 1 pc (comoving) \citep{wise14}.
This suggests that the prediction of the escape fraction and star formation rates 
is likely to be converged in (sub-)parsec-scale simulations, as the physics of the 
collapse and subsequent disruption of star-forming clouds is reasonably well captured.

\section{Conclusions}

Using the radiation-hydrodynamic simulation code, {\sc ramses-rt}, we investigate the 
importance of mini-haloes with $10^{6.25} \la \mhalo \la 10^{8}\,\msun$ 
for  the reionisation history of the Universe. For this purpose, we run nine zoom-in, 
cosmological simulations with non-equilibrium photo-chemistry involving molecular 
hydrogen, star formation based on the local conditions of the ISM 
(i.e. thermo-turbulent model), radiation pressure from UV and IR photons, 
heating by photo-ionisation, and mechanical SN feedback. We measure the stellar mass 
and escape fractions from the simulations at $7\le z < 18$, and examine the 
relative importance of  mini-haloes for  reionisation 
by computing the evolution of the fraction of ionised hydrogen 
based on a simple analytic model. Our conclusions can be summarised as follows.

\begin{itemize}
\item[(i)] We find that even though the instantaneous escape fraction is highly variable, 
the photon-number weighted average escape fraction is generally high ($\sim20$--$40\%$) 
in  mini-haloes (Figure~\ref{fig:fescavg}), in agreement with previous 
studies \citep{wise14,xu16}. The escape fraction is nevertheless occasionally  
low even after a burst of star formation, and we attribute this to the fact that 
in this case the disruption of the birth clouds is too slow for LyC photons to escape efficiently (Figure~\ref{fig:pdf_low}). \\ 

\item[(ii)] The process primarily  responsible for the efficient escape of LyC photons 
in  mini-haloes is heating due to photoionisaton (Figure~\ref{fig:mech}). 
Direct radiation pressure alone seems unlikely to create low-density channels 
through which ionising radiation can escape.  Rather, it simply compresses gas radially,
resulting in dense neutral gas shells (Figure~\ref{fig:physics}). 
Because star formation is very stochastic with our  thermo-turbulent model, 
strong radiation from young stars plays a significant role before SN explosions come into play.
This leads to a short time-delay of $\la 5\,{\rm Myr}$ between the peak of the 
production rate of LyC photons and the escape fraction when ionising radiation 
can efficiently escape into the IGM (Figures~\ref{fig:fesc} and \ref{fig:twopeak}).\\

\item[(iii)] Only massive Pop III stars ($M_{\rm PopIII}\ga100\,\msun$) can play 
a role in ionising the neutral gas beyond the virial radius. LyC photons from Pop III stars 
with lower masses ($M_{\rm PopIII}\la70\,\msun$) are mostly absorbed 
in the central region of their host halo (Figure~\ref{fig:fesc_pop3}). \\

\item[(iv)] The escape fraction for individual star formation events is not strongly 
correlated with the star formation rate (Figure~\ref{fig:why}). 
Instead, we find the the production rate of ionising photons per dense gas mass is the key
to determining the escape fraction. Mini-haloes producing a large number of LyC 
photons per dense gas mass show a higher escape fraction. 
This also explains why the escape of LyC photons is significant in Pop III stars 
with $M_{\rm PopIII}\ga100\,\msun$. \\

\item[(v)] Star formation is efficiently regulated by feedback processes in the mini-haloes. 
The typical stellar mass  in the mini-halo ranges from $10^3$ -- a few times 
$10^4\,\msun$, meaning that only less than a few percent of baryons is converted 
into stars (Figure~\ref{fig:mstar}).  Our simulated galaxies follow the normalisation and slope 
of the $\mstar-\mhalo$ sequence obtained from the radiation-hydrodynamic simulations 
by \citet{kimm14}, which is interestingly similar to the $z\sim0$ empirical sequence 
by \citet{behroozi13} when extrapolated to the mini-halo regime. \\

\item[(vi)] Using a simple analytic approach, we find that the main driver of reionisation
is the LyC photons from atomic-cooling haloes ($10^8\,\msun \la \mhalo \la 10^{11}\,\msun$).
Although mini-haloes are more abundant and their escape fraction is higher than 
that of the more massive haloes, their contribution to reionisation is of minor importance,
essentially because star formation is inefficient (Figure~\ref{fig:tau}). Even if 100\% of 
the mini-haloes are assumed to host stars, this does not  alter our conclusions. 
Our fiducial model with a reasonably high escape fraction in the atomic-cooling 
regime predicts that the universe is fully ionised by $z\sim6.7$ and $\tau_e=0.067$.
Although the optical depth estimation is consistent with the latest Planck measurement 
within the two sigma uncertainty, reionisation is predicted to end relatively early, 
which may be in tension with the rather late end of reionization suggested 
by the most recent CMB data as well as the \lya\ forest and \lya\ emitter data. 
A larger escape of  $\sim$  20\%  in massive haloes with  $10^{9}-10^{11}\,\msun$ 
would further raise this tension. Further investigation of the escape fraction in  atomic-cooling haloes 
should shed light on this potential over-production of LyC photons.
\end{itemize}

\section*{Acknowledgements}
We thank  the anonymous referee for helpful comments that improved the manuscript.
We are grateful to John Wise and Renyue Cen for helpful conversations and Romain Teyssier 
for making his code {\sc ramses} publicly available. 
We also thank John Wise and Hao Xu for sharing the occupation fraction and the
Pop III star formation rate data with us. 
This work was supported by the ERC Advanced Grant 320596
``The Emergence of Structure during the Epoch of Reionization" 
and the Spin(e) grants ANR-13-BS05-0005 of the French Agence Nationale de la Recherche (http://cosmicorigin.org).
HK is supported by Foundation Boustany, the Isaac Newton Studentship, and the Cambridge Overseas Trust.
JR was funded by the European Research Council under the European Unions Seventh Framework Programme (FP7/2007- 2013)/ERC Grant agreement 278594-GasAroundGalaxies, and the Marie Curie Training Network CosmoComp (PITN- GA-2009-238356).
JD and AS's research is supported by funding from Adrian Beecroft, the Oxford Martin School and the STFC.
This work used the DiRAC Complexity system, operated by the University of Leicester IT Services, which forms part of the STFC DiRAC HPC Facility (www.dirac.ac.uk ).   This equipment is funded by BIS National E-Infrastructure capital grant ST/K000373/1 and  STFC DiRAC Operations grant ST/K0003259/1. DiRAC is part of the National E-Infrastructure.

\small
\bibliographystyle{mnras}
\bibliography{refs}

\end{document}